\PassOptionsToPackage{dvipsnames}{xcolor}
\documentclass[manuscript=false,authorversion,review=false,nonacm]{acmart}\usepackage[]{graphicx}\usepackage[]{xcolor}
\makeatletter
\def\maxwidth{ %
  \ifdim\Gin@nat@width>\linewidth
    \linewidth
  \else
    \Gin@nat@width
  \fi
}
\makeatother

\definecolor{fgcolor}{rgb}{0.345, 0.345, 0.345}
\newcommand{\hlkwb}[1]{\textcolor[rgb]{0.69,0.353,0.396}{#1}}%

\usepackage{framed}
\makeatletter
 {\par\unskip\endMakeFramed%
 \at@end@of@kframe}
\makeatother

\definecolor{shadecolor}{rgb}{.97, .97, .97}
\definecolor{messagecolor}{rgb}{0, 0, 0}
\definecolor{warningcolor}{rgb}{1, 0, 1}
\definecolor{errorcolor}{rgb}{1, 0, 0}

\usepackage{alltt}

\usepackage{epsfig,endnotes,url}

\usepackage{xspace}

\usepackage{amsmath, amsthm, latexsym} 
\usepackage{ulem}

\usepackage{graphicx}

\usepackage{multirow}
\usepackage{booktabs}
\usepackage{longtable}
\usepackage{rotating}
\usepackage{array}

\usepackage{wasysym}

\usepackage{siunitx}
\usepackage{pdfpages}

\usepackage{paralist}

\usepackage{balance}
\usepackage{adjustbox}

\usepackage{versions}
\excludeversion{DVSupplementaryAppendix}
\includeversion{DVSupplementaryExternal}

\usepackage[caption=false]{subfig}

\usepackage{hyperref}

\hypersetup{
bookmarks=true,
colorlinks=false,
pdftitle={Validity and Reliability of IUIPC-8}
}
\hypersetup{draft}

\newcommand{\defterm}[2]{\hypertarget{#1}{\textit{#2}}}
\newcommand{\refterm}[2]{\hyperlink{#1}{#2}}

\theoremstyle{definition} 

\newcommand{\vari}[1]{\ensuremath{\mathit{#1}\xspace}}
\newcommand{\const}[1]{\ensuremath{\mathsf{#1}\xspace}}

\newcommand{\fName}[1]{#1\xspace}
\newcommand{\fVar}[1]{\ensuremath{\const{#1}}}
\newcommand{\iVar}[1]{\ensuremath{\const{#1}}}

\newcommand{\CASCAde}{ERC Starting Grant CASCAde (GA n\textsuperscript{o}716980)}
\IfFileExists{upquote.sty}{\usepackage{upquote}}{}

\renewcommand\footnotetextcopyrightpermission[1]{}
\setcopyright{none}
\begin{document}

















\newcommand{\inputCorSDBV}{
\begin{table*}[p]
\centering\caption{Correlations and standard deviations of samples \textsf{B} and \textsf{V} used in the study (before outlier treatment)}
\label{tab:inputCorSDBV}
\captionsetup{position=top}
\subfloat[Sample \textsf{B}]{
\label{tab:inputCorSDBraw}
\centering
\begingroup\footnotesize
\begin{tabular}{rrrrrrrrrrr}
  \toprule
 & 1 & 2 & 3 & 4 & 5 & 6 & 7 & 8 & 9 & 10 \\ 
  \midrule
1. ctrl1 &  &  &  &  &  &  &  &  &  &  \\ 
  2. ctrl2 & 0.617 &  &  &  &  &  &  &  &  &  \\ 
  3. ctrl3 & 0.238 & 0.274 &  &  &  &  &  &  &  &  \\ 
  4. awa1 & 0.230 & 0.261 & 0.292 &  &  &  &  &  &  &  \\ 
  5. awa2 & 0.265 & 0.308 & 0.321 & 0.653 &  &  &  &  &  &  \\ 
  6. awa3 & 0.224 & 0.187 & 0.256 & 0.311 & 0.286 &  &  &  &  &  \\ 
  7. coll1 & 0.017 & 0.028 & 0.263 & 0.062 & 0.106 & 0.329 &  &  &  &  \\ 
  8. coll2 & 0.071 & 0.062 & 0.296 & 0.240 & 0.251 & 0.274 & 0.650 &  &  &  \\ 
  9. coll3 & 0.130 & 0.079 & 0.309 & 0.139 & 0.188 & 0.399 & 0.765 & 0.692 &  &  \\ 
  10. coll4 & 0.165 & 0.113 & 0.315 & 0.268 & 0.244 & 0.481 & 0.717 & 0.613 & 0.810 &  \\ 
  SD & 1.018 & 1.024 & 1.008 & 0.593 & 0.581 & 0.852 & 1.380 & 1.122 & 1.290 & 1.278 \\ 
   \bottomrule
\multicolumn{11}{c}{\emph{Note:} $N_\mathsf{B} = 379$}\\
\end{tabular}
\endgroup
}

\subfloat[Sample \textsf{V}]{%
\label{tab:inputCorSDVraw}
\centering
\begingroup\footnotesize
\begin{tabular}{rrrrrrrrrrr}
  \toprule
 & 1 & 2 & 3 & 4 & 5 & 6 & 7 & 8 & 9 & 10 \\ 
  \midrule
1. ctrl1 &  &  &  &  &  &  &  &  &  &  \\ 
  2. ctrl2 & 0.614 &  &  &  &  &  &  &  &  &  \\ 
  3. ctrl3 & 0.276 & 0.270 &  &  &  &  &  &  &  &  \\ 
  4. awa1 & 0.278 & 0.291 & 0.385 &  &  &  &  &  &  &  \\ 
  5. awa2 & 0.263 & 0.257 & 0.350 & 0.589 &  &  &  &  &  &  \\ 
  6. awa3 & 0.171 & 0.210 & 0.248 & 0.326 & 0.389 &  &  &  &  &  \\ 
  7. coll1 & 0.027 & 0.075 & 0.249 & 0.127 & 0.190 & 0.281 &  &  &  &  \\ 
  8. coll2 & 0.137 & 0.162 & 0.270 & 0.161 & 0.227 & 0.346 & 0.531 &  &  &  \\ 
  9. coll3 & 0.117 & 0.102 & 0.303 & 0.221 & 0.270 & 0.348 & 0.713 & 0.599 &  &  \\ 
  10. coll4 & 0.085 & 0.137 & 0.319 & 0.263 & 0.287 & 0.330 & 0.595 & 0.511 & 0.757 &  \\ 
  SD & 1.112 & 1.100 & 1.056 & 0.756 & 0.721 & 1.063 & 1.400 & 1.063 & 1.231 & 1.217 \\ 
   \bottomrule
\multicolumn{11}{c}{\emph{Note:} $N_\mathsf{V} = 433$}\\
\end{tabular}
\endgroup
}
\end{table*}
}

\newcommand{\histograms}{
\begin{figure}[p]
\centering\caption{Histograms and density plots of all samples}
\captionsetup{position=top}
\label{fig:histograms}
\subfloat[Sample \textsf{A}]{
\label{fig:histA}
\centering\includegraphics[keepaspectratio,width=.8\maxwidth]{./figure/histA.pdf}}

\subfloat[Sample \textsf{B}]{
\label{fig:histB}
\centering\includegraphics[keepaspectratio,width=.8\maxwidth]{./figure/histB.pdf}}

\subfloat[Sample \textsf{V}]{%
\label{fig:ihistV}
\centering\includegraphics[keepaspectratio,width=.8\maxwidth]{./figure/histV.pdf}}
\end{figure}
}
\newcommand{\inputCorSDB}{
\begin{table*}[ht]
\centering
\caption{Correlations and standard deviations of Sample \textsf{B}} 
\label{tab:inputCorSDB}
\begingroup\footnotesize
\begin{tabular}{rrrrrrrrrrr}
  \toprule
 & 1 & 2 & 3 & 4 & 5 & 6 & 7 & 8 & 9 & 10 \\ 
  \midrule
1. ctrl1 &  &  &  &  &  &  &  &  &  &  \\ 
  2. ctrl2 & 0.56 &  &  &  &  &  &  &  &  &  \\ 
  3. ctrl3 & 0.25 & 0.27 &  &  &  &  &  &  &  &  \\ 
  4. awa1 & 0.25 & 0.23 & 0.25 &  &  &  &  &  &  &  \\ 
  5. awa2 & 0.32 & 0.32 & 0.30 & 0.62 &  &  &  &  &  &  \\ 
  6. awa3 & 0.23 & 0.19 & 0.26 & 0.32 & 0.31 &  &  &  &  &  \\ 
  7. coll1 & 0.05 & 0.05 & 0.26 & 0.05 & 0.11 & 0.33 &  &  &  &  \\ 
  8. coll2 & 0.10 & 0.06 & 0.28 & 0.22 & 0.24 & 0.30 & 0.66 &  &  &  \\ 
  9. coll3 & 0.16 & 0.09 & 0.32 & 0.16 & 0.22 & 0.40 & 0.76 & 0.72 &  &  \\ 
  10. coll4 & 0.19 & 0.10 & 0.31 & 0.23 & 0.23 & 0.47 & 0.71 & 0.62 & 0.81 &  \\ 
  SD & 0.93 & 0.93 & 1.00 & 0.55 & 0.56 & 0.82 & 1.36 & 1.11 & 1.24 & 1.22 \\ 
   \bottomrule
\multicolumn{11}{c}{\emph{Note:} $N_\mathsf{B} = 370$}\\
\end{tabular}
\endgroup
\end{table*}
}
\newcommand{\inputCorSDMB}{
\begin{table*}[ht]
\centering
\caption{Correlations, means and standard deviations of base Sample \textsf{B}} 
\label{tab:inputCorSDMB}
\begingroup\footnotesize
\begin{tabular}{rrrrrrrrrrr}
  \toprule
 & 1 & 2 & 3 & 4 & 5 & 6 & 7 & 8 & 9 & 10 \\ 
  \midrule
1. ctrl1 &  &  &  &  &  &  &  &  &  &  \\ 
  2. ctrl2 & 0.56 &  &  &  &  &  &  &  &  &  \\ 
  3. ctrl3 & 0.25 & 0.27 &  &  &  &  &  &  &  &  \\ 
  4. awa1 & 0.25 & 0.23 & 0.25 &  &  &  &  &  &  &  \\ 
  5. awa2 & 0.32 & 0.32 & 0.30 & 0.62 &  &  &  &  &  &  \\ 
  6. awa3 & 0.23 & 0.19 & 0.26 & 0.32 & 0.31 &  &  &  &  &  \\ 
  7. coll1 & 0.05 & 0.05 & 0.26 & 0.05 & 0.11 & 0.33 &  &  &  &  \\ 
  8. coll2 & 0.10 & 0.06 & 0.28 & 0.22 & 0.24 & 0.30 & 0.66 &  &  &  \\ 
  9. coll3 & 0.16 & 0.09 & 0.32 & 0.16 & 0.22 & 0.40 & 0.76 & 0.72 &  &  \\ 
  10. coll4 & 0.19 & 0.10 & 0.31 & 0.23 & 0.23 & 0.47 & 0.71 & 0.62 & 0.81 &  \\ 
  M & 5.97 & 5.96 & 6.68 & 6.62 & 5.26 & 5.76 & 5.69 & 5.73 & 5.97 & 5.96 \\ 
  SD & 0.93 & 0.93 & 1.00 & 0.55 & 0.56 & 0.82 & 1.36 & 1.11 & 1.24 & 1.22 \\ 
   \bottomrule
\multicolumn{11}{c}{\emph{Note:} $N_\mathsf{B} = 370$}\\
\end{tabular}
\endgroup
\end{table*}
}
\newcommand{\inputCorSDMV}{
\begin{table*}[ht]
\centering
\caption{Correlations, means and standard deviations of validation Sample \textsf{V}} 
\label{tab:inputCorSDMV}
\begingroup\footnotesize
\begin{tabular}{rrrrrrrrrrr}
  \toprule
 & 1 & 2 & 3 & 4 & 5 & 6 & 7 & 8 & 9 & 10 \\ 
  \midrule
1. ctrl1 &  &  &  &  &  &  &  &  &  &  \\ 
  2. ctrl2 & 0.53 &  &  &  &  &  &  &  &  &  \\ 
  3. ctrl3 & 0.29 & 0.31 &  &  &  &  &  &  &  &  \\ 
  4. awa1 & 0.28 & 0.34 & 0.32 &  &  &  &  &  &  &  \\ 
  5. awa2 & 0.27 & 0.31 & 0.29 & 0.55 &  &  &  &  &  &  \\ 
  6. awa3 & 0.15 & 0.23 & 0.18 & 0.21 & 0.27 &  &  &  &  &  \\ 
  7. coll1 & 0.06 & 0.13 & 0.25 & 0.12 & 0.21 & 0.30 &  &  &  &  \\ 
  8. coll2 & 0.18 & 0.22 & 0.27 & 0.13 & 0.22 & 0.37 & 0.52 &  &  &  \\ 
  9. coll3 & 0.14 & 0.14 & 0.30 & 0.18 & 0.26 & 0.34 & 0.71 & 0.58 &  &  \\ 
  10. coll4 & 0.07 & 0.15 & 0.31 & 0.23 & 0.26 & 0.29 & 0.59 & 0.49 & 0.74 &  \\ 
  M & 5.94 & 5.90 & 6.59 & 6.61 & 5.25 & 5.78 & 5.73 & 5.75 & 5.94 & 5.90 \\ 
  SD & 1.02 & 1.00 & 1.03 & 0.69 & 0.62 & 0.98 & 1.39 & 1.04 & 1.19 & 1.18 \\ 
   \bottomrule
\multicolumn{11}{c}{\emph{Note:} $N_\mathsf{V} = 419$}\\
\end{tabular}
\endgroup
\end{table*}
}
\newcommand{\inputCorSD}{
\begin{table*}[p]
\centering\caption{Correlations and standard deviations of samples used in the study}
\captionsetup{position=top}
\label{tab:inputCorSD}
\subfloat[Sample \textsf{A}]{
\label{tab:inputCorSDA}
\centering
\begingroup\footnotesize
\begin{tabular}{rrrrrrrrrrr}
  \toprule
 & 1 & 2 & 3 & 4 & 5 & 6 & 7 & 8 & 9 & 10 \\ 
  \midrule
1. ctrl1 &  &  &  &  &  &  &  &  &  &  \\ 
  2. ctrl2 & 0.531 &  &  &  &  &  &  &  &  &  \\ 
  3. ctrl3 & 0.298 & 0.420 &  &  &  &  &  &  &  &  \\ 
  4. awa1 & 0.238 & 0.456 & 0.393 &  &  &  &  &  &  &  \\ 
  5. awa2 & 0.313 & 0.367 & 0.306 & 0.494 &  &  &  &  &  &  \\ 
  6. awa3 & 0.306 & 0.282 & 0.142 & 0.130 & 0.062 &  &  &  &  &  \\ 
  7. coll1 & 0.094 & 0.072 & 0.092 & 0.162 & 0.128 & 0.268 &  &  &  &  \\ 
  8. coll2 & 0.129 & 0.134 & 0.236 & 0.194 & 0.225 & 0.273 & 0.567 &  &  &  \\ 
  9. coll3 & 0.105 & 0.184 & 0.236 & 0.304 & 0.214 & 0.170 & 0.674 & 0.608 &  &  \\ 
  10. coll4 & 0.167 & 0.182 & 0.261 & 0.309 & 0.292 & 0.313 & 0.580 & 0.393 & 0.660 &  \\ 
  SD & 1.031 & 1.216 & 1.201 & 0.942 & 0.871 & 1.097 & 1.357 & 1.273 & 1.371 & 1.310 \\ 
   \bottomrule
\multicolumn{11}{c}{\emph{Note:} $N_{\mathsf{A}}^\prime = 201$}\\
\end{tabular}
\endgroup
}

\subfloat[Sample \textsf{B}]{
\label{tab:inputCorSDB}
\centering
\begingroup\footnotesize
\begin{tabular}{rrrrrrrrrrr}
  \toprule
 & 1 & 2 & 3 & 4 & 5 & 6 & 7 & 8 & 9 & 10 \\ 
  \midrule
1. ctrl1 &  &  &  &  &  &  &  &  &  &  \\ 
  2. ctrl2 & 0.557 &  &  &  &  &  &  &  &  &  \\ 
  3. ctrl3 & 0.253 & 0.275 &  &  &  &  &  &  &  &  \\ 
  4. awa1 & 0.253 & 0.232 & 0.253 &  &  &  &  &  &  &  \\ 
  5. awa2 & 0.316 & 0.325 & 0.296 & 0.619 &  &  &  &  &  &  \\ 
  6. awa3 & 0.226 & 0.192 & 0.262 & 0.321 & 0.312 &  &  &  &  &  \\ 
  7. coll1 & 0.053 & 0.050 & 0.264 & 0.050 & 0.107 & 0.325 &  &  &  &  \\ 
  8. coll2 & 0.102 & 0.058 & 0.281 & 0.219 & 0.236 & 0.297 & 0.656 &  &  &  \\ 
  9. coll3 & 0.157 & 0.087 & 0.321 & 0.163 & 0.220 & 0.401 & 0.756 & 0.717 &  &  \\ 
  10. coll4 & 0.194 & 0.099 & 0.310 & 0.234 & 0.234 & 0.470 & 0.706 & 0.618 & 0.812 &  \\ 
  SD & 0.931 & 0.933 & 0.996 & 0.549 & 0.563 & 0.819 & 1.358 & 1.106 & 1.242 & 1.218 \\ 
   \bottomrule
\multicolumn{11}{c}{\emph{Note:} $N_{\mathsf{B}}^\prime = 370$}\\
\end{tabular}
\endgroup
}

\subfloat[Sample \textsf{V}]{%
\label{tab:inputCorSDV}
\centering
\begingroup\footnotesize
\begin{tabular}{rrrrrrrrrrr}
  \toprule
 & 1 & 2 & 3 & 4 & 5 & 6 & 7 & 8 & 9 & 10 \\ 
  \midrule
1. ctrl1 &  &  &  &  &  &  &  &  &  &  \\ 
  2. ctrl2 & 0.534 &  &  &  &  &  &  &  &  &  \\ 
  3. ctrl3 & 0.288 & 0.309 &  &  &  &  &  &  &  &  \\ 
  4. awa1 & 0.277 & 0.339 & 0.320 &  &  &  &  &  &  &  \\ 
  5. awa2 & 0.272 & 0.310 & 0.291 & 0.549 &  &  &  &  &  &  \\ 
  6. awa3 & 0.154 & 0.232 & 0.176 & 0.210 & 0.274 &  &  &  &  &  \\ 
  7. coll1 & 0.065 & 0.126 & 0.255 & 0.120 & 0.214 & 0.295 &  &  &  &  \\ 
  8. coll2 & 0.175 & 0.218 & 0.272 & 0.127 & 0.218 & 0.366 & 0.522 &  &  &  \\ 
  9. coll3 & 0.144 & 0.143 & 0.296 & 0.183 & 0.265 & 0.339 & 0.705 & 0.581 &  &  \\ 
  10. coll4 & 0.075 & 0.148 & 0.305 & 0.226 & 0.257 & 0.294 & 0.591 & 0.493 & 0.743 &  \\ 
  SD & 1.017 & 0.998 & 1.031 & 0.691 & 0.622 & 0.982 & 1.390 & 1.036 & 1.192 & 1.179 \\ 
   \bottomrule
\multicolumn{11}{c}{\emph{Note:} $N_{\mathsf{V}}^\prime = 419$}\\
\end{tabular}
\endgroup
}
\end{table*}
}
\newcommand{\descSubScales}{
\begin{table}[htb]
\centering\caption{Means (SDs) of the summarized sub-scales of IUIPC-10}
\label{tab:descSubScales}
\begingroup\footnotesize
\begin{tabular}{rll}
  \toprule
 & Sample \textsf{B} & Sample \textsf{V} \\ 
  \midrule
\textsf{ctrl} & $5.93$ $(0.78)$ & $5.86$ $(0.84)$ \\ 
  \textsf{awa} & $6.51$ $(0.52)$ & $6.43$ $(0.66)$ \\ 
  \textsf{coll} & $5.58$ $(1.12)$ & $5.60$ $(1.04)$ \\ 
  \textsf{iuipc} & $6.00$ $(0.61)$ & $5.96$ $(0.64)$ \\ 
   \bottomrule
\end{tabular}
\endgroup
\end{table}
}

\newcommand{\densitySubScalesRedux}{
\definecolor{virililac}{HTML}{101474}
\definecolor{viriviolet}{HTML}{351042}
\definecolor{virigreen}{HTML}{317F79}
\definecolor{viriyellow}{HTML}{FCE528}
\definecolor{viriorange}{HTML}{F4AC27}
\begin{figure*}[tb]
\centering\captionsetup{position=bottom}
\begin{minipage}{0.24\textwidth}%
\subfloat[Control]{
\label{fig:densityCtrl}
\centering
\includegraphics[width=\maxwidth]{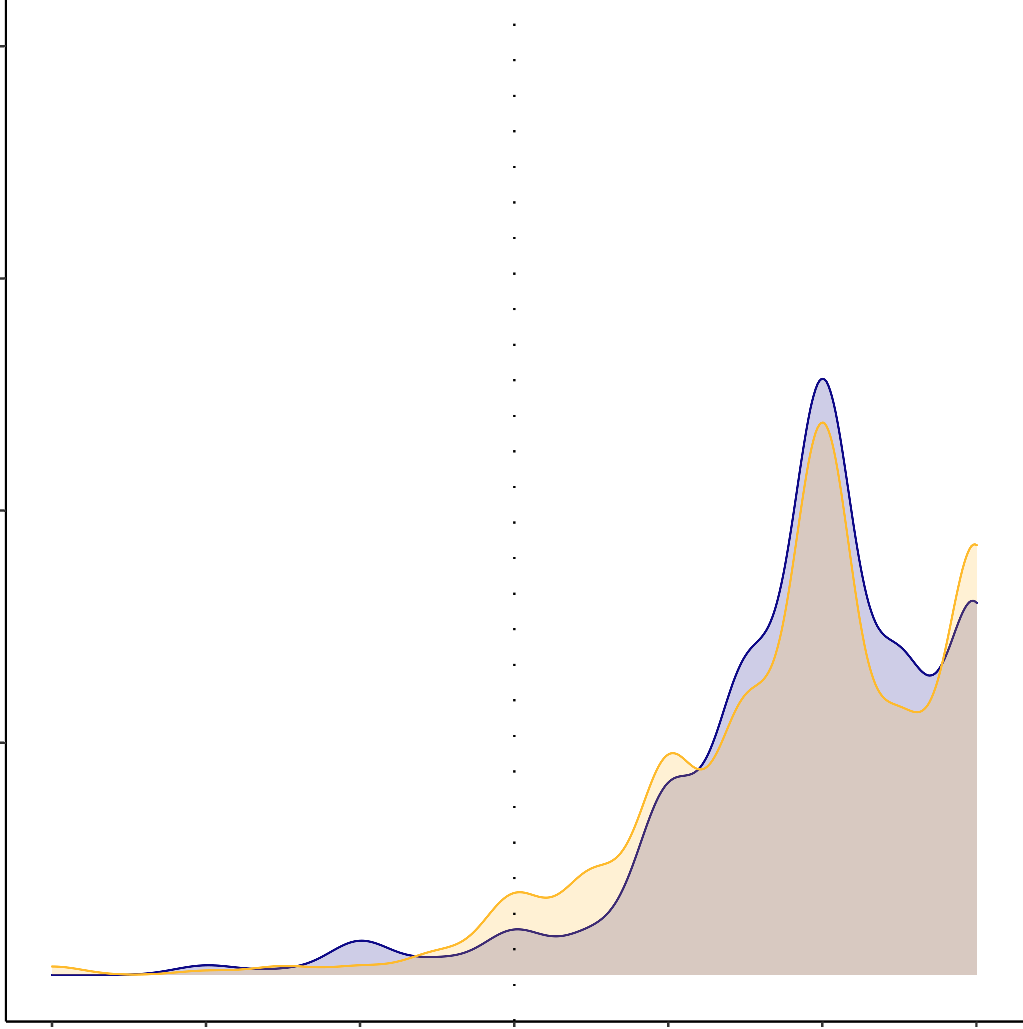} 
}
\end{minipage}~
\begin{minipage}{0.24\textwidth}%
\subfloat[Awareness]{
\label{fig:densityAwa}
\centering
\includegraphics[width=\maxwidth]{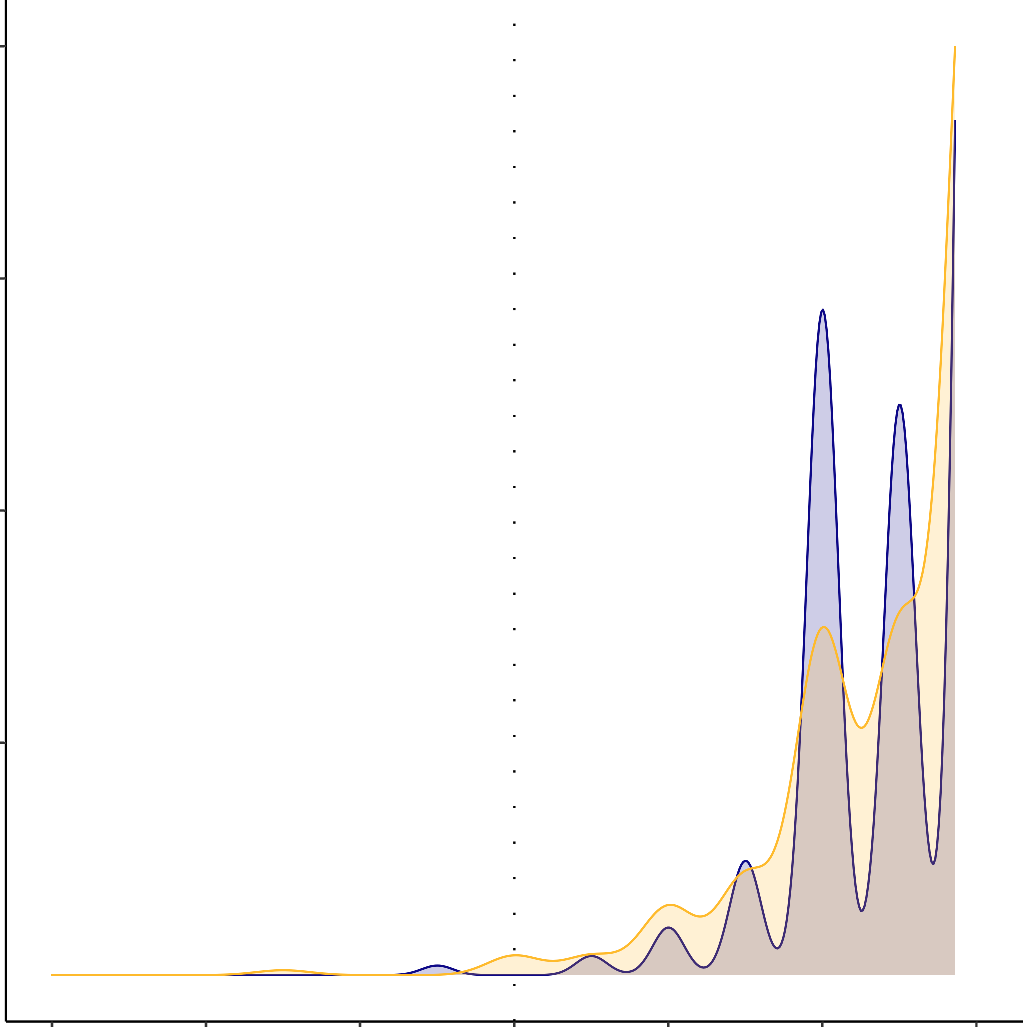} 
}
\end{minipage}~
\begin{minipage}{0.24\textwidth}%
\subfloat[Collection]{%
\label{fig:densityColl}
\centering
\includegraphics[width=\maxwidth]{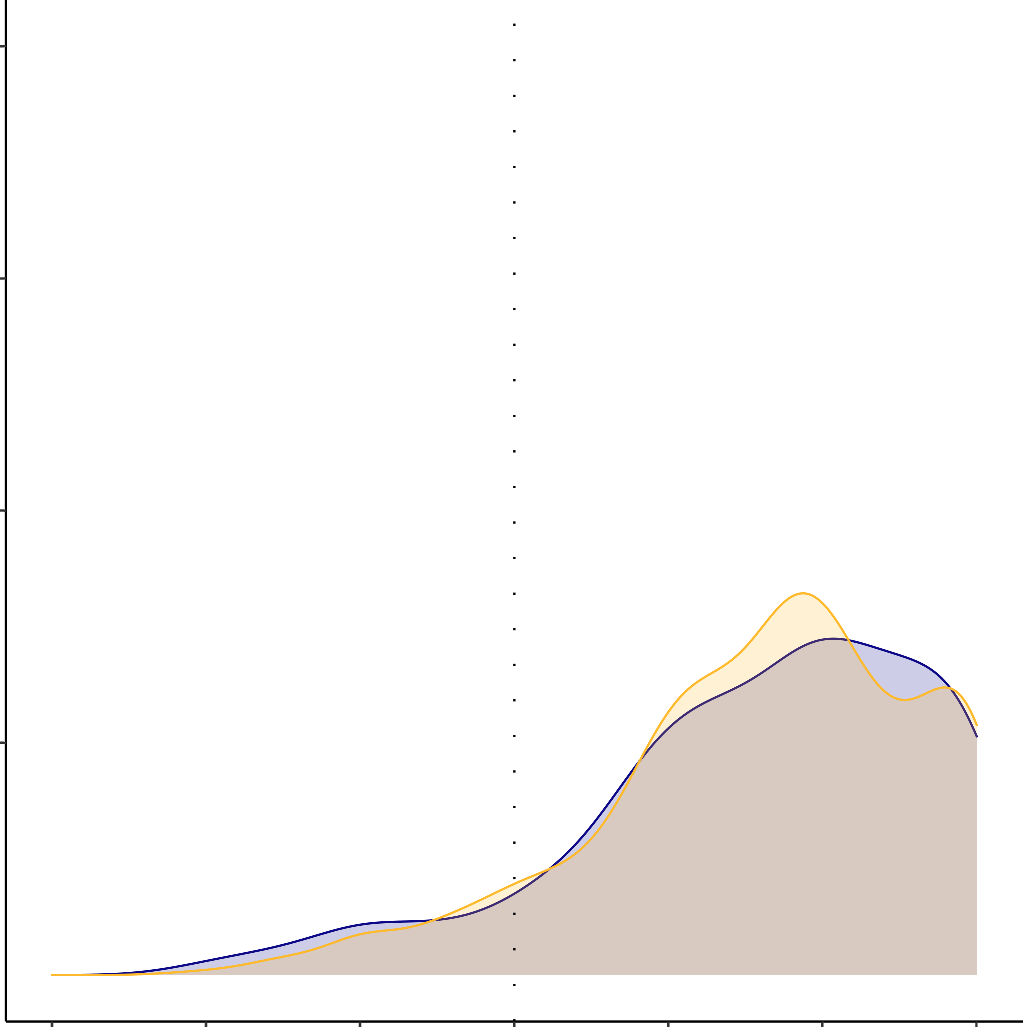} 
}
\end{minipage}
\begin{minipage}{0.24\textwidth}%
\subfloat[IUIPC-8 Overall]{%
\label{fig:densityIUIPC}
\centering
\includegraphics[width=\maxwidth]{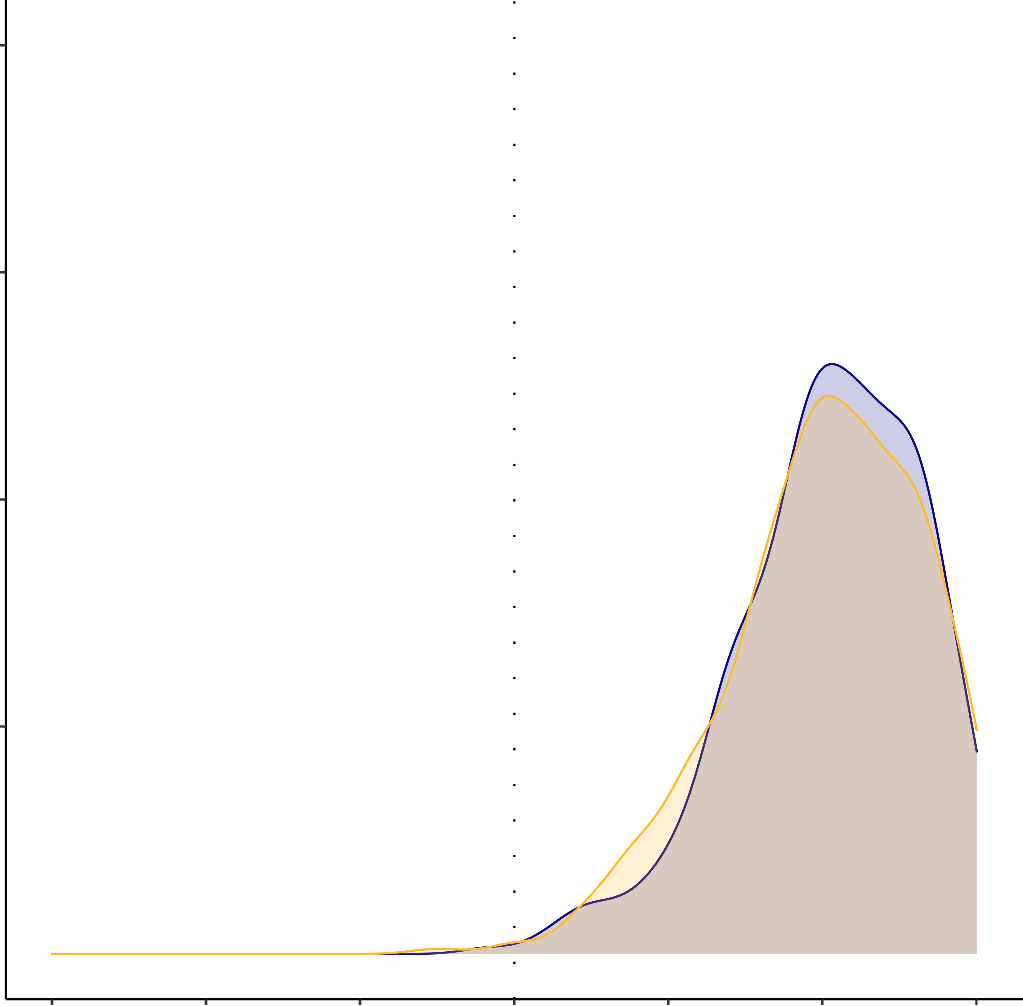} 
}
\end{minipage}
\caption{Density of IUIPC-8 subscale responses across samples (\textsf{B}: \textcolor{virililac}{violet}, \textsf{V}: \textcolor{viriorange}{orange}). \emph{Note:} All graphs are on the same scale}
\label{fig:densitySubScalesRedux}
\Description[Density plots to show the distributions of the sub-scales ctrl, aware, and collect and the average of all items for both samples]{The figure shows four density plots each containing the distribution of base sample B and validation sample V. The first three plots contain the density plots for the sub-scales ctrl, aware, and collect. The final graph shows the flat average of all items. The graphs show that the the distributions are negatively skewed.}
\end{figure*}
}

\newcommand{\demoPyramid}{
\definecolor{viriviolet}{HTML}{351042}
\definecolor{virigreen}{HTML}{317F79}
\definecolor{viriyellow}{HTML}{FCE528}
\begin{figure*}[tb]
\centering\captionsetup{position=bottom}
\begin{minipage}{0.45\textwidth}%
\subfloat[Sample \textsf{B}]{
\label{fig:pyramidB}
\centering
\includegraphics[width=\maxwidth]{figure/demo_table-1} 
}
\end{minipage}
\begin{minipage}{0.45\textwidth}%
\subfloat[Sample \textsf{V}]{
\label{fig:pyramidV}
\centering
\includegraphics[width=\maxwidth]{figure/demo_table-2} 
}
\end{minipage}
\caption{Sample demographics distribution by age and gender (\textsf{female}: \textcolor{viriviolet}{violet}, \textsf{male}: \textcolor{viriyellow}{yellow}).}
\label{fig:demoPyramid}
\end{figure*}
}

\newcommand{\residualsBwls}{
\begin{table*}[tbp]
\centering\caption{Residuals of the WLSMV-estimated CFA of IUIPC-10 on Sample \textsf{B}}
\label{tab:residualsBwls}
\captionsetup{position=top}
\subfloat[Correlation residuals]{
\label{tab:residualsBwlscor}
\centering
\begingroup\footnotesize
\begin{tabular}{rllllllllll}
  \toprule
 & 1 & 2 & 3 & 4 & 5 & 6 & 7 & 8 & 9 & 10 \\ 
  \midrule
1. ctrl1 & --- &  &  &  &  &  &  &  &  &  \\ 
  2. ctrl2 & \textbf{0.123} & --- &  &  &  &  &  &  &  &  \\ 
  3. ctrl3 & \textbf{-0.151} & \textbf{-0.111} & --- &  &  &  &  &  &  &  \\ 
  4. awa1 & -0.003 & -0.017 & 0.015 & --- &  &  &  &  &  &  \\ 
  5. awa2 & 0.043 & 0.052 & 0.03 & 0.09 & --- &  &  &  &  &  \\ 
  6. awa3 & -0.054 & -0.075 & 0.008 & \textbf{-0.151} & \textbf{-0.223} & --- &  &  &  &  \\ 
  7. coll1 & \textbf{-0.151} & \textbf{-0.163} & 0.082 & \textbf{-0.248} & \textbf{-0.214} & 0.079 & --- &  &  &  \\ 
  8. coll2 & -0.084 & \textbf{-0.129} & \textbf{0.151} & -0.041 & -0.032 & 0.078 & 0.039 & --- &  &  \\ 
  9. coll3 & -0.053 & \textbf{-0.118} & \textbf{0.133} & \textbf{-0.152} & \textbf{-0.111} & \textbf{0.129} & -0.003 & 0.005 & --- &  \\ 
  10. coll4 & -0.007 & -0.088 & \textbf{0.138} & -0.049 & -0.076 & \textbf{0.221} & 0.008 & -0.043 & 0 & --- \\ 
   \bottomrule
\multicolumn{11}{c}{\emph{Note:} Correlation residuals in absolute $> 0.1$ are marked}\\
\end{tabular}
\endgroup
}

\subfloat[Covariance residuals]{%
\label{tab:residualsBwlsstd}
\centering
\begingroup\footnotesize
\begin{tabular}{rllllllllll}
  \toprule
 & 1 & 2 & 3 & 4 & 5 & 6 & 7 & 8 & 9 & 10 \\ 
  \midrule
1. ctrl1 & --- &  &  &  &  &  &  &  &  &  \\ 
  2. ctrl2 & 0.123 & --- &  &  &  &  &  &  &  &  \\ 
  3. ctrl3 & -0.151 & -0.111 & --- &  &  &  &  &  &  &  \\ 
  4. awa1 & -0.003 & -0.017 & 0.015 & --- &  &  &  &  &  &  \\ 
  5. awa2 & 0.043 & 0.052 & 0.03 & 0.09 & --- &  &  &  &  &  \\ 
  6. awa3 & -0.054 & -0.075 & 0.008 & -0.151 & -0.223 & --- &  &  &  &  \\ 
  7. coll1 & -0.151 & -0.163 & 0.082 & -0.248 & -0.214 & 0.079 & --- &  &  &  \\ 
  8. coll2 & -0.084 & -0.129 & 0.151 & -0.041 & -0.032 & 0.078 & 0.039 & --- &  &  \\ 
  9. coll3 & -0.053 & -0.118 & 0.133 & -0.152 & -0.111 & 0.129 & -0.003 & 0.005 & --- &  \\ 
  10. coll4 & -0.007 & -0.088 & 0.138 & -0.049 & -0.076 & 0.221 & 0.008 & -0.043 & 0 & --- \\ 
   \bottomrule
\multicolumn{11}{c}{\emph{Note:} Standardized residuals are not available for estimator WLSMV}\\
\end{tabular}
\endgroup
}
\end{table*}
}
\newcommand{\corfitBreduxwls}{
\begin{table}[ht]
\centering
\caption{Implied correlations of the improved CFA model on Sample \textsf{B}} 
\label{tab:corfitBreduxwls}
\begingroup\footnotesize
\begin{tabular}{rlll}
  \toprule
 & 1 & 2 & 3 \\ 
  \midrule
1. ctrl & 0.646 &  &  \\ 
  2. aware & 0.573 & 0.802 &  \\ 
  3. collect & 0.221 & 0.316 & 0.772 \\ 
   \bottomrule
\multicolumn{4}{c}{\emph{Note:} The diagonal contains the $\sqrt{\mathit{AVE}}$}\\
\end{tabular}
\endgroup
\end{table}
}
\newcommand{\residualsBreduxwls}{
\begin{table*}[p]
\centering\caption{Residuals of the WLSMV  CFA of IUIPC-8 on Sample \textsf{B}}
\label{tab:residualsBreduxwls}
\captionsetup{position=top}
\subfloat[Correlation residuals]{
\label{tab:residualsBreduxwlscor}
\centering
\begingroup\footnotesize
\begin{tabular}{rllllllll}
  \toprule
 & 1 & 2 & 3 & 4 & 5 & 6 & 7 & 8 \\ 
  \midrule
1. ctrl1 & --- &  &  &  &  &  &  &  \\ 
  2. ctrl2 & 0 & --- &  &  &  &  &  &  \\ 
  3. awa1 & -0.017 & 0.003 & --- &  &  &  &  &  \\ 
  4. awa2 & -0.017 & 0.032 & 0 & --- &  &  &  &  \\ 
  5. coll1 & -0.071 & -0.068 & \textbf{-0.132} & \textbf{-0.115} & --- &  &  &  \\ 
  6. coll2 & -0.005 & -0.037 & 0.071 & 0.065 & 0.034 & --- &  &  \\ 
  7. coll3 & 0.04 & -0.009 & -0.019 & 0.003 & -0.012 & 0.003 & --- &  \\ 
  8. coll4 & 0.082 & 0.015 & 0.078 & 0.034 & 0.008 & -0.036 & 0.004 & --- \\ 
   \bottomrule
\multicolumn{9}{c}{\emph{Note:} Correlation residuals in absolute $> 0.1$ are marked}\\
\end{tabular}
\endgroup
}

\subfloat[Covariance residuals]{%
\label{tab:residualsBreduxwlsraw}
\centering
\begingroup\footnotesize
\begin{tabular}{rllllllll}
  \toprule
 & 1 & 2 & 3 & 4 & 5 & 6 & 7 & 8 \\ 
  \midrule
1. ctrl1 & --- &  &  &  &  &  &  &  \\ 
  2. ctrl2 & 0 & --- &  &  &  &  &  &  \\ 
  3. awa1 & -0.017 & 0.003 & --- &  &  &  &  &  \\ 
  4. awa2 & -0.017 & 0.032 & 0 & --- &  &  &  &  \\ 
  5. coll1 & -0.071 & -0.068 & -0.132 & -0.115 & --- &  &  &  \\ 
  6. coll2 & -0.005 & -0.037 & 0.071 & 0.065 & 0.034 & --- &  &  \\ 
  7. coll3 & 0.04 & -0.009 & -0.019 & 0.003 & -0.012 & 0.003 & --- &  \\ 
  8. coll4 & 0.082 & 0.015 & 0.078 & 0.034 & 0.008 & -0.036 & 0.004 & --- \\ 
   \bottomrule
\multicolumn{9}{c}{\emph{Note:} Standardized residuals are not available for estimator WLSMV}\\
\end{tabular}
\endgroup
}
\end{table*}
}
\newcommand{\loadingsBreduxwls}{
\begin{table*}[tbp]
\centering
\caption{Factor loadings and their standardized solution of the WLSMV CFA of IUIPC-8 on Sample~\textsf{B}} 
\label{tab:loadingsBreduxwls}
\begingroup\footnotesize
\begin{tabular}{lllrrrrrrrrrrrr}
  \toprule
 \multirow{2}{*}{Factor} & \multirow{2}{*}{Indicator} & \multicolumn{4}{c}{Factor Loading} & \multicolumn{4}{c}{Standardized Solution} & \multicolumn{5}{c}{Reliability} \\ 
 \cmidrule(lr){3-6} \cmidrule(lr){7-10} \cmidrule(lr){11-15} 
  & & $\lambda$ & $\vari{SE}_{\lambda}$ & $Z_{\lambda}$ & $p_{\lambda}$ & $\beta$ & $\vari{SE}_{\beta}$ & $Z_{\beta}$ & $p_{\beta}$ & $R^2$ & $\mathit{AVE}$ & $\alpha$ & $\omega$ & $\mathit{S/\!N}_\omega$ \\
  \midrule
 ctrl & ctrl1 & $1.00^+$ &  &  &  & 0.84 & 0.06 & 14.21 & $<.001$ & 0.70 & 0.65 & 0.72 & 0.72 & 2.57 \\ 
   & ctrl2 & $0.91\phantom{^+}$ & 0.13 & 7.22 & $<.001$ & 0.77 & 0.06 & 13.18 & $<.001$ & 0.59 &  &  &  &  \\ 
  aware & awa1 & $1.00^+$ &  &  &  & 0.82 & 0.06 & 14.07 & $<.001$ & 0.66 & 0.80 & 0.76 & 0.79 & 3.67 \\ 
   & awa2 & $1.19\phantom{^+}$ & 0.15 & 7.91 & $<.001$ & 0.97 & 0.06 & 16.41 & $<.001$ & 0.94 &  &  &  &  \\ 
  collect & coll1 & $1.00^+$ &  &  &  & 0.85 & 0.02 & 50.30 & $<.001$ & 0.72 & 0.77 & 0.91 & 0.91 & 10.18 \\ 
   & coll2 & $0.95\phantom{^+}$ & 0.03 & 37.56 & $<.001$ & 0.81 & 0.02 & 40.59 & $<.001$ & 0.65 &  &  &  &  \\ 
   & coll3 & $1.14\phantom{^+}$ & 0.02 & 49.25 & $<.001$ & 0.97 & 0.01 & 102.10 & $<.001$ & 0.93 &  &  &  &  \\ 
   & coll4 & $1.05\phantom{^+}$ & 0.02 & 44.24 & $<.001$ & 0.89 & 0.01 & 60.34 & $<.001$ & 0.79 &  &  &  &  \\ 
  iuipc & collect & $0.30\phantom{^+}$ & 0.05 & 5.56 & $<.001$ & 0.35 & 0.06 & 5.57 & $<.001$ & 0.12 &  &  &  &  \\ 
   & ctrl & $0.53\phantom{^+}$ & 0.08 & 6.62 & $<.001$ & 0.63 & 0.10 & 6.54 & $<.001$ & 0.40 &  &  &  &  \\ 
   & aware & $0.74\phantom{^+}$ & 0.12 & 6.00 & $<.001$ & 0.91 & 0.12 & 7.27 & $<.001$ & 0.82 &  &  &  &  \\ 
   \bottomrule
\multicolumn{14}{c}{\emph{Note:} $^+$ fixed parameter; the standardized solution is STDALL}\\
\end{tabular}
\endgroup
\end{table*}
}
\newcommand{\CorVarDevB}{
\begin{table}[ht]
\centering
\caption{Sample correlations and correlation residuals of the fitted model on Sample \textsf{B}} 
\label{tab:CorVarDevB}
\begingroup\footnotesize
\begin{tabular}{rrrrrrrrr}
  \toprule
 & ctrl1 & ctrl2 & awa1 & awa2 & coll1 & coll2 & coll3 & coll4 \\ 
  \midrule
ctrl1 & 0.866 & -0.000 & -0.017 & -0.071 & 0.082 & -0.037 & 0.071 & 0.034 \\ 
  ctrl2 & 0.557 & 0.871 & -0.017 & -0.005 & 0.003 & -0.009 & -0.019 & 0.034 \\ 
  awa1 & 0.253 & 0.232 & 0.301 & 0.040 & 0.032 & 0.015 & 0.078 & -0.012 \\ 
  awa2 & 0.316 & 0.325 & 0.619 & 0.317 & -0.068 & -0.000 & -0.115 & 0.008 \\ 
  coll1 & 0.053 & 0.050 & 0.050 & 0.107 & 1.843 & -0.132 & 0.065 & 0.003 \\ 
  coll2 & 0.102 & 0.058 & 0.219 & 0.236 & 0.656 & 1.224 & 0.003 & -0.036 \\ 
  coll3 & 0.157 & 0.087 & 0.163 & 0.220 & 0.756 & 0.717 & 1.542 & 0.004 \\ 
  coll4 & 0.194 & 0.099 & 0.234 & 0.234 & 0.706 & 0.618 & 0.812 & 1.482 \\ 
   \bottomrule
\end{tabular}
\endgroup
\end{table}
}
\newcommand{\thresholdsB}{
\begin{table}[ht]
\centering
\caption{Thresholds table of fitted model on Sample \textsf{B}} 
\label{tab:thresholdsB}
\begingroup\footnotesize
\begin{tabular}{rlrrrrrr}
  \toprule
 & Variable & t1 & t2 & t3 & t4 & t5 & t6 \\ 
  \midrule
1 & ctrl1 & -2.782 & -1.926 & -1.583 & -0.705 & 0.524 &  \\ 
  2 & ctrl2 & -1.846 & -1.537 & -0.722 & 0.564 &  &  \\ 
  3 & awa1 & -2.782 & -1.810 & -0.564 &  &  &  \\ 
  4 & awa2 & -2.549 & -1.885 & -0.404 &  &  &  \\ 
  5 & coll1 & -2.782 & -1.776 & -1.115 & -0.731 & 0.075 & 0.842 \\ 
  6 & coll2 & -2.297 & -1.632 & -1.251 & -0.382 & 0.588 &  \\ 
  7 & coll3 & -1.972 & -1.416 & -1.128 & -0.310 & 0.517 &  \\ 
  8 & coll4 & -2.782 & -2.139 & -1.515 & -1.066 & -0.382 & 0.501 \\ 
   \bottomrule
\end{tabular}
\endgroup
\end{table}
}

\newcommand{\loadingsVwls}{
\begin{table*}[tbp]
\centering
\caption{Factor loadings and their standardized solution of the WLSMV CFA of IUIPC-10 on validation Sample~\textsf{V}} 
\label{tab:loadingsVwls}
\begingroup\footnotesize
\begin{tabular}{lllrrrrrrrrrrrr}
  \toprule
 \multirow{2}{*}{Factor} & \multirow{2}{*}{Indicator} & \multicolumn{4}{c}{Factor Loading} & \multicolumn{4}{c}{Standardized Solution} & \multicolumn{5}{c}{Reliability} \\ 
 \cmidrule(lr){3-6} \cmidrule(lr){7-10} \cmidrule(lr){11-15} 
  & & $\lambda$ & $\vari{SE}_{\lambda}$ & $Z_{\lambda}$ & $p_{\lambda}$ & $\beta$ & $\vari{SE}_{\beta}$ & $Z_{\beta}$ & $p_{\beta}$ & $R^2$ & $\mathit{AVE}$ & $\alpha$ & $\omega$ & $\mathit{S/\!N}_\omega$ \\
  \midrule
 ctrl & ctrl1 & $1.00^+$ &  &  &  & 0.74 & 0.03 & 24.31 & $<.001$ & 0.54 & 0.55 & 0.64 & 0.74 & 2.79 \\ 
   & ctrl2 & $1.09\phantom{^+}$ & 0.08 & 14.29 & $<.001$ & 0.81 & 0.03 & 25.02 & $<.001$ & 0.65 &  &  &  &  \\ 
   & ctrl3 & $0.92\phantom{^+}$ & 0.07 & 13.64 & $<.001$ & 0.67 & 0.04 & 17.12 & $<.001$ & 0.45 &  &  &  &  \\ 
  aware & awa1 & $1.00^+$ &  &  &  & 0.77 & 0.03 & 24.09 & $<.001$ & 0.59 & 0.56 & 0.56 & 0.67 & 2.00 \\ 
   & awa2 & $1.12\phantom{^+}$ & 0.08 & 14.34 & $<.001$ & 0.86 & 0.04 & 24.42 & $<.001$ & 0.73 &  &  &  &  \\ 
   & awa3 & $0.80\phantom{^+}$ & 0.06 & 12.73 & $<.001$ & 0.61 & 0.04 & 15.22 & $<.001$ & 0.37 &  &  &  &  \\ 
  collect & coll1 & $1.00^+$ &  &  &  & 0.80 & 0.02 & 38.82 & $<.001$ & 0.64 & 0.69 & 0.86 & 0.87 & 6.94 \\ 
   & coll2 & $0.90\phantom{^+}$ & 0.04 & 23.71 & $<.001$ & 0.72 & 0.02 & 29.11 & $<.001$ & 0.52 &  &  &  &  \\ 
   & coll3 & $1.20\phantom{^+}$ & 0.04 & 32.35 & $<.001$ & 0.95 & 0.01 & 79.36 & $<.001$ & 0.91 &  &  &  &  \\ 
   & coll4 & $1.06\phantom{^+}$ & 0.03 & 36.43 & $<.001$ & 0.84 & 0.02 & 51.11 & $<.001$ & 0.71 &  &  &  &  \\ 
  iuipc & collect & $0.43\phantom{^+}$ & 0.04 & 11.68 & $<.001$ & 0.54 & 0.04 & 12.38 & $<.001$ & 0.30 &  &  &  &  \\ 
   & ctrl & $0.53\phantom{^+}$ & 0.04 & 12.32 & $<.001$ & 0.72 & 0.05 & 15.80 & $<.001$ & 0.53 &  &  &  &  \\ 
   & aware & $0.74\phantom{^+}$ & 0.06 & 12.99 & $<.001$ & 0.96 & 0.06 & 15.97 & $<.001$ & 0.93 &  &  &  &  \\ 
   \bottomrule
\multicolumn{14}{c}{\emph{Note:} $^+$ fixed parameter; the standardized solution is STDALL}\\
\end{tabular}
\endgroup
\end{table*}
}
\newcommand{\loadingsVreduxwls}{
\begin{table*}[tbp]
\centering
\caption{Factor loadings and their standardized solution of the WLSMV CFA of IUIPC-8 on Sample~\textsf{V}} 
\label{tab:loadingsVreduxwls}
\begingroup\footnotesize
\begin{tabular}{lllrrrrrrrrrrrr}
  \toprule
 \multirow{2}{*}{Factor} & \multirow{2}{*}{Indicator} & \multicolumn{4}{c}{Factor Loading} & \multicolumn{4}{c}{Standardized Solution} & \multicolumn{5}{c}{Reliability} \\ 
 \cmidrule(lr){3-6} \cmidrule(lr){7-10} \cmidrule(lr){11-15} 
  & & $\lambda$ & $\vari{SE}_{\lambda}$ & $Z_{\lambda}$ & $p_{\lambda}$ & $\beta$ & $\vari{SE}_{\beta}$ & $Z_{\beta}$ & $p_{\beta}$ & $R^2$ & $\mathit{AVE}$ & $\alpha$ & $\omega$ & $\mathit{S/\!N}_\omega$ \\
  \midrule
 ctrl & ctrl1 & $1.00^+$ &  &  &  & 0.77 & 0.04 & 19.57 & $<.001$ & 0.59 & 0.66 & 0.70 & 0.74 & 2.86 \\ 
   & ctrl2 & $1.12\phantom{^+}$ & 0.11 & 10.34 & $<.001$ & 0.86 & 0.04 & 19.53 & $<.001$ & 0.74 &  &  &  &  \\ 
  aware & awa1 & $1.00^+$ &  &  &  & 0.79 & 0.04 & 20.52 & $<.001$ & 0.63 & 0.72 & 0.71 & 0.73 & 2.64 \\ 
   & awa2 & $1.14\phantom{^+}$ & 0.10 & 11.23 & $<.001$ & 0.90 & 0.04 & 20.65 & $<.001$ & 0.81 &  &  &  &  \\ 
  collect & coll1 & $1.00^+$ &  &  &  & 0.80 & 0.02 & 39.41 & $<.001$ & 0.64 & 0.69 & 0.86 & 0.87 & 6.95 \\ 
   & coll2 & $0.89\phantom{^+}$ & 0.04 & 23.66 & $<.001$ & 0.71 & 0.02 & 28.73 & $<.001$ & 0.50 &  &  &  &  \\ 
   & coll3 & $1.20\phantom{^+}$ & 0.04 & 32.92 & $<.001$ & 0.96 & 0.01 & 82.08 & $<.001$ & 0.92 &  &  &  &  \\ 
   & coll4 & $1.05\phantom{^+}$ & 0.03 & 37.36 & $<.001$ & 0.84 & 0.02 & 52.27 & $<.001$ & 0.70 &  &  &  &  \\ 
  iuipc & collect & $0.34\phantom{^+}$ & 0.05 & 7.42 & $<.001$ & 0.42 & 0.06 & 7.55 & $<.001$ & 0.18 &  &  &  &  \\ 
   & ctrl & $0.49\phantom{^+}$ & 0.06 & 8.23 & $<.001$ & 0.64 & 0.07 & 9.75 & $<.001$ & 0.41 &  &  &  &  \\ 
   & aware & $0.78\phantom{^+}$ & 0.09 & 8.78 & $<.001$ & 0.99 & 0.10 & 10.40 & $<.001$ & 0.98 &  &  &  &  \\ 
   \bottomrule
\multicolumn{14}{c}{\emph{Note:} $^+$ fixed parameter; the standardized solution is STDALL}\\
\end{tabular}
\endgroup
\end{table*}
}
\newcommand{\residualsVreduxwls}{
\begin{table*}[p]
\centering\caption{Residuals of the WLSMV validation CFA of IUIPC-8 on Sample \textsf{V}}
\label{tab:residualsVreduxwls}
\captionsetup{position=top}
\subfloat[Correlation residuals]{
\label{tab:residualsVreduxcorwls}
\centering
\begingroup\footnotesize
\begin{tabular}{rllllllll}
  \toprule
 & 1 & 2 & 3 & 4 & 5 & 6 & 7 & 8 \\ 
  \midrule
1. ctrl1 & --- &  &  &  &  &  &  &  \\ 
  2. ctrl2 & 0 & --- &  &  &  &  &  &  \\ 
  3. awa1 & 0.036 & 0.026 & --- &  &  &  &  &  \\ 
  4. awa2 & -0.015 & -0.027 & 0 & --- &  &  &  &  \\ 
  5. coll1 & -0.069 & -0.035 & -0.064 & -0.001 & --- &  &  &  \\ 
  6. coll2 & 0.083 & 0.086 & -0.036 & 0.057 & 0.023 & --- &  &  \\ 
  7. coll3 & -0.002 & -0.021 & -0.029 & 0.002 & 0.005 & -0.019 & --- &  \\ 
  8. coll4 & -0.045 & 0.012 & 0.043 & 0.045 & -0.012 & -0.014 & 0.007 & --- \\ 
   \bottomrule
\multicolumn{9}{c}{\emph{Note:} Correlation residuals in absolute $> 0.1$ are marked}\\
\end{tabular}
\endgroup
}

\subfloat[Covariance residuals]{%
\label{tab:residualsVreduxrawwls}
\centering
\begingroup\footnotesize
\begin{tabular}{rllllllll}
  \toprule
 & 1 & 2 & 3 & 4 & 5 & 6 & 7 & 8 \\ 
  \midrule
1. ctrl1 & --- &  &  &  &  &  &  &  \\ 
  2. ctrl2 & 0 & --- &  &  &  &  &  &  \\ 
  3. awa1 & 0.036 & 0.026 & --- &  &  &  &  &  \\ 
  4. awa2 & -0.015 & -0.027 & 0 & --- &  &  &  &  \\ 
  5. coll1 & -0.069 & -0.035 & -0.064 & -0.001 & --- &  &  &  \\ 
  6. coll2 & 0.083 & 0.086 & -0.036 & 0.057 & 0.023 & --- &  &  \\ 
  7. coll3 & -0.002 & -0.021 & -0.029 & 0.002 & 0.005 & -0.019 & --- &  \\ 
  8. coll4 & -0.045 & 0.012 & 0.043 & 0.045 & -0.012 & -0.014 & 0.007 & --- \\ 
   \bottomrule
\multicolumn{9}{c}{\emph{Note:} Standardized residuals are not available for estimator WLSMV}\\
\end{tabular}
\endgroup
}
\end{table*}
}
\newcommand{\residualsVwls}{
\begin{table*}[p]
\centering\caption{Residuals of the WLSMV validation CFA of IUIPC-10 on Sample \textsf{V}}
\label{tab:residualsVwls}
\captionsetup{position=top}
\subfloat[Correlation residuals]{
\label{tab:residualsVwlscor}
\centering
\begingroup\footnotesize
\begin{tabular}{rllllllllll}
  \toprule
 & 1 & 2 & 3 & 4 & 5 & 6 & 7 & 8 & 9 & 10 \\ 
  \midrule
1. ctrl1 & --- &  &  &  &  &  &  &  &  &  \\ 
  2. ctrl2 & 0.066 & --- &  &  &  &  &  &  &  &  \\ 
  3. ctrl3 & \textbf{-0.108} & \textbf{-0.155} & --- &  &  &  &  &  &  &  \\ 
  4. awa1 & 0.024 & 0.023 & 0.086 & --- &  &  &  &  &  &  \\ 
  5. awa2 & -0.021 & -0.022 & 0.021 & 0.055 & --- &  &  &  &  &  \\ 
  6. awa3 & -0.067 & -0.027 & -0.028 & \textbf{-0.104} & -0.068 & --- &  &  &  &  \\ 
  7. coll1 & \textbf{-0.137} & \textbf{-0.105} & 0.092 & \textbf{-0.123} & -0.06 & 0.098 & --- &  &  &  \\ 
  8. coll2 & 0.02 & 0.021 & \textbf{0.142} & -0.092 & -0.001 & \textbf{0.178} & 0.017 & --- &  &  \\ 
  9. coll3 & -0.082 & \textbf{-0.104} & 0.097 & -0.097 & -0.068 & \textbf{0.103} & 0.011 & -0.024 & --- &  \\ 
  10. coll4 & \textbf{-0.116} & -0.062 & \textbf{0.139} & -0.02 & -0.019 & 0.084 & -0.012 & -0.023 & 0.01 & --- \\ 
   \bottomrule
\multicolumn{11}{c}{\emph{Note:} Correlation residuals in absolute $> 0.1$ are marked}\\
\end{tabular}
\endgroup
}

\subfloat[Covariance residuals]{%
\label{tab:residualsVwlsraw}
\centering
\begingroup\footnotesize
\begin{tabular}{rllllllllll}
  \toprule
 & 1 & 2 & 3 & 4 & 5 & 6 & 7 & 8 & 9 & 10 \\ 
  \midrule
1. ctrl1 & --- &  &  &  &  &  &  &  &  &  \\ 
  2. ctrl2 & 0.066 & --- &  &  &  &  &  &  &  &  \\ 
  3. ctrl3 & -0.108 & -0.155 & --- &  &  &  &  &  &  &  \\ 
  4. awa1 & 0.024 & 0.023 & 0.086 & --- &  &  &  &  &  &  \\ 
  5. awa2 & -0.021 & -0.022 & 0.021 & 0.055 & --- &  &  &  &  &  \\ 
  6. awa3 & -0.067 & -0.027 & -0.028 & -0.104 & -0.068 & --- &  &  &  &  \\ 
  7. coll1 & -0.137 & -0.105 & 0.092 & -0.123 & -0.06 & 0.098 & --- &  &  &  \\ 
  8. coll2 & 0.02 & 0.021 & 0.142 & -0.092 & -0.001 & 0.178 & 0.017 & --- &  &  \\ 
  9. coll3 & -0.082 & -0.104 & 0.097 & -0.097 & -0.068 & 0.103 & 0.011 & -0.024 & --- &  \\ 
  10. coll4 & -0.116 & -0.062 & 0.139 & -0.02 & -0.019 & 0.084 & -0.012 & -0.023 & 0.01 & --- \\ 
   \bottomrule
\multicolumn{11}{c}{\emph{Note:} Standardized residuals are not available for estimator WLSMV}\\
\end{tabular}
\endgroup
}
\end{table*}
}
\newcommand{\corfitV}{
\begin{table}[ht]
\centering
\caption{Implied correlations of the CFA model on Sample \textsf{V}} 
\label{tab:corfitV}
\begingroup\footnotesize
\begin{tabular}{rlll}
  \toprule
 & 1 & 2 & 3 \\ 
  \midrule
1. ctrl & 0.549 &  &  \\ 
  2. aware & 0.698 & 0.564 &  \\ 
  3. collect & 0.395 & 0.524 & 0.691 \\ 
   \bottomrule
\multicolumn{4}{c}{\emph{Note:} The diagonal contains the $\sqrt{\mathit{AVE}}$}\\
\end{tabular}
\endgroup
\end{table}
}
\newcommand{\corfitVredux}{
\begin{table}[ht]
\centering
\caption{Implied correlations of the improved CFA model on Sample \textsf{V}} 
\label{tab:corfitVredux}
\begingroup\footnotesize
\begin{tabular}{rlll}
  \toprule
 & 1 & 2 & 3 \\ 
  \midrule
1. ctrl & 0.663 &  &  \\ 
  2. aware & 0.63 & 0.718 &  \\ 
  3. collect & 0.268 & 0.415 & 0.69 \\ 
   \bottomrule
\multicolumn{4}{c}{\emph{Note:} The diagonal contains the $\sqrt{\mathit{AVE}}$}\\
\end{tabular}
\endgroup
\end{table}
}
\newcommand{\CorVarDevV}{
\begin{table}[ht]
\centering
\caption{Sample correlations and correlation residuals of the fitted model on Sample \textsf{V}} 
\label{tab:CorVarDevV}
\begingroup\footnotesize
\begin{tabular}{rrrrrrrrr}
  \toprule
 & ctrl1 & ctrl2 & awa1 & awa2 & coll1 & coll2 & coll3 & coll4 \\ 
  \midrule
ctrl1 & 0.866 & 0.000 & 0.036 & -0.069 & -0.045 & 0.086 & -0.036 & 0.045 \\ 
  ctrl2 & 0.557 & 0.871 & -0.015 & 0.083 & 0.026 & -0.021 & -0.029 & 0.023 \\ 
  awa1 & 0.253 & 0.232 & 0.301 & -0.002 & -0.027 & 0.012 & 0.043 & 0.005 \\ 
  awa2 & 0.316 & 0.325 & 0.619 & 0.317 & -0.035 & -0.000 & -0.001 & -0.012 \\ 
  coll1 & 0.053 & 0.050 & 0.050 & 0.107 & 1.843 & -0.064 & 0.057 & -0.019 \\ 
  coll2 & 0.102 & 0.058 & 0.219 & 0.236 & 0.656 & 1.224 & 0.002 & -0.014 \\ 
  coll3 & 0.157 & 0.087 & 0.163 & 0.220 & 0.756 & 0.717 & 1.542 & 0.007 \\ 
  coll4 & 0.194 & 0.099 & 0.234 & 0.234 & 0.706 & 0.618 & 0.812 & 1.482 \\ 
   \bottomrule
\end{tabular}
\endgroup
\end{table}
}
\newcommand{\thresholdsV}{
\begin{table}[ht]
\centering
\caption{Thresholds table of fitted model on Sample \textsf{V}} 
\label{tab:thresholdsV}
\begingroup\footnotesize
\begin{tabular}{rlrrrrrr}
  \toprule
 & Variable & t1 & t2 & t3 & t4 & t5 & t6 \\ 
  \midrule
1 & ctrl1 & -2.822 & -2.592 & -1.939 & -1.382 & -0.607 & 0.448 \\ 
  2 & ctrl2 & -2.449 & -1.833 & -1.464 & -0.592 & 0.543 &  \\ 
  3 & awa1 & -2.822 & -2.188 & -1.397 & -0.488 &  &  \\ 
  4 & awa2 & -2.259 & -1.644 & -0.442 &  &  &  \\ 
  5 & coll1 & -2.592 & -1.744 & -1.087 & -0.650 & 0.021 & 0.857 \\ 
  6 & coll2 & -2.822 & -2.344 & -1.717 & -1.397 & -0.442 & 0.695 \\ 
  7 & coll3 & -2.073 & -1.447 & -1.228 & -0.345 & 0.543 &  \\ 
  8 & coll4 & -2.259 & -1.599 & -1.109 & -0.332 & 0.468 &  \\ 
   \bottomrule
\end{tabular}
\endgroup
\end{table}
}

\newcommand{\instumentReliabilityTable}{
\begin{table*}[ht]
\centering
\caption{Item/construct reliability metrics of IUIPC-10 on Sample \textsf{B}} 
\label{tab:instruments.reliability}
\begingroup\footnotesize
\begin{tabular}{lrrrrrrlrr}
  \toprule
Construct & $\alpha$ & $\mathsf{LL}(\alpha)$ & $\mathsf{UL}(\alpha)$ & $\lambda_{6}$ & $\mathsf{avg}(r)$ & $r_{\mathsf{i/w}}$ & Bartlett $p$ & \textsf{KMO} & \textsf{MSA} \\ 
  \midrule
Control & 0.63 & 0.56 & 0.69 & 0.56 & 0.36 &  & $<.001$ & 0.59 &  \\ 
  \qquad ctrl1 &  &  &  &  &  & 0.79 &  &  & 0.56 \\ 
  \qquad ctrl2 &  &  &  &  &  & 0.80 &  &  & 0.56 \\ 
  \qquad ctrl3 &  &  &  &  &  & 0.69 &  &  & 0.76 \\ 
  Awareness & 0.68 & 0.62 & 0.73 & 0.62 & 0.42 &  & $<.001$ & 0.60 &  \\ 
  \qquad awa1 &  &  &  &  &  & 0.77 &  &  & 0.57 \\ 
  \qquad awa2 &  &  &  &  &  & 0.77 &  &  & 0.57 \\ 
  \qquad awa3 &  &  &  &  &  & 0.78 &  &  & 0.79 \\ 
  Collection & 0.91 & 0.89 & 0.92 & 0.89 & 0.71 &  & $<.001$ & 0.83 &  \\ 
  \qquad coll1 &  &  &  &  &  & 0.89 &  &  & 0.88 \\ 
  \qquad coll2 &  &  &  &  &  & 0.83 &  &  & 0.88 \\ 
  \qquad coll3 &  &  &  &  &  & 0.93 &  &  & 0.77 \\ 
  \qquad coll4 &  &  &  &  &  & 0.89 &  &  & 0.81 \\ 
   \bottomrule
\end{tabular}
\endgroup
\end{table*}
}

\newcommand{\descSubScalesOrig}{
\begin{table}[htb]
\centering\caption{Means (SDs) of the summarized sub-scales of IUIPC-10}
\label{tab:descSubScalesOrig}
\begin{adjustbox}{max width=\columnwidth}
\begingroup\footnotesize
\begin{tabular}{rlll}
  \toprule
 & Sample \textsf{B} & Sample \textsf{V} & Malhotra et al.~\cite{malhotra2004internet} \\ 
  \midrule
\textsf{ctrl} & $5.93$ $(0.78)$ & $5.86$ $(0.84)$ & $5.67$ $(1.06)$ \\ 
  \textsf{awa} & $6.51$ $(0.52)$ & $6.43$ $(0.66)$ & $6.21$ $(0.87)$ \\ 
  \textsf{coll} & $5.58$ $(1.12)$ & $5.60$ $(1.04)$ & $5.63$ $(1.09)$ \\ 
  \textsf{iuipc} & $6.00$ $(0.61)$ & $5.96$ $(0.64)$ & $5.84$ $(1.01)$ \\ 
   \bottomrule
\end{tabular}
\endgroup
\end{adjustbox}\end{table}
}
\newcommand{\descSubScalesRedux}{
\begin{table}[htb]
\centering\caption{Means (SDs) of the summarized sub-scales of IUIPC-8 and the original IUIPC-10}
\label{tab:descSubScalesRedux}
\begin{adjustbox}{max width=\columnwidth}
\begingroup\footnotesize
\begin{tabular}{rlll}
  \toprule
 & Sample \textsf{B} & Sample \textsf{V} & Malhotra et al.~\cite{malhotra2004internet} \\ 
  \midrule
\textsf{ctrl} & $5.92$ $(0.92)$ & $5.86$ $(0.99)$ & $5.67$ $(1.06)$ \\ 
  \textsf{aware} & $6.64$ $(0.53)$ & $6.56$ $(0.66)$ & $6.21$ $(0.87)$ \\ 
  \textsf{collect} & $5.58$ $(1.12)$ & $5.60$ $(1.04)$ & $5.63$ $(1.09)$ \\ 
  \textsf{iuipc} & $6.05$ $(0.60)$ & $6.01$ $(0.63)$ & $5.84$ $(1.01)$ \\ 
   \bottomrule
\end{tabular}
\endgroup
\end{adjustbox}\\\emph{Note:} \textsf{iuipc} is the flat mean of all items of the scale.\end{table}
}

\newcommand{\discriminantBredux}{
\begin{table*}[p]
\centering\caption{Evidence for discriminant validity of IUIPC-8 (Fornell-Larcker Criterion and Heterotrait-Monotrait Ratio) on Sample \textsf{B}}
\captionsetup{position=top}
\label{tab:discriminantBredux}
\subfloat[Fornell-Larcker]{
\label{tab:corfitBredux}
\centering
\begingroup\footnotesize
\begin{tabular}{rlll}
  \toprule
 & 1 & 2 & 3 \\ 
  \midrule
1. ctrl & 0.646 &  &  \\ 
  2. aware & 0.573 & 0.802 &  \\ 
  3. collect & 0.221 & 0.316 & 0.772 \\ 
   \bottomrule
\multicolumn{4}{c}{\emph{Note:} The diagonal contains the $\sqrt{\mathit{AVE}}$}\\
\end{tabular}
\endgroup
}~\subfloat[HTMT Ratio]{%
\label{tab:htmtBredux}
\centering
\begingroup\footnotesize
\begin{tabular}{rlll}
  \toprule
 & 1 & 2 & 3 \\ 
  \midrule
1. ctrl & --- &  &  \\ 
  2. aware & 0.47 & --- &  \\ 
  3. collect & 0.14 & 0.25 & --- \\ 
   \bottomrule
\end{tabular}
\endgroup
}
\end{table*}
}
\newcommand{\discriminantVredux}{
\begin{table*}[p]
\centering\caption{Evidence for discriminant validity of IUIPC-8 (Fornell-Larcker Criterion and Heterotrait-Monotrait Ratio) on Sample \textsf{V}}
\captionsetup{position=top}
\label{tab:discriminantVredux}
\subfloat[Fornell-Larcker]{
\label{tab:corfitVredux}
\centering
\begingroup\footnotesize
\begin{tabular}{rlll}
  \toprule
 & 1 & 2 & 3 \\ 
  \midrule
1. ctrl & 0.663 &  &  \\ 
  2. aware & 0.63 & 0.718 &  \\ 
  3. collect & 0.268 & 0.415 & 0.69 \\ 
   \bottomrule
\multicolumn{4}{c}{\emph{Note:} The diagonal contains the $\sqrt{\mathit{AVE}}$}\\
\end{tabular}
\endgroup
}~\subfloat[HTMT Ratio]{%
\label{tab:htmtVredux}
\centering
\begingroup\footnotesize
\begin{tabular}{rlll}
  \toprule
 & 1 & 2 & 3 \\ 
  \midrule
1. ctrl & --- &  &  \\ 
  2. aware & 0.55 & --- &  \\ 
  3. collect & 0.23 & 0.34 & --- \\ 
   \bottomrule
\end{tabular}
\endgroup
}
\end{table*}
}
\newcommand{\corfitBVredux}{
\begin{table*}[p]
\centering\caption{First-level correlations and $\sqrt{\vari{AVE}}$ as evidence for the Fornell-Larcker criterion for discriminant validity on IUIPC-8}
\captionsetup{position=top}
\label{tab:corfitBVredux}
\subfloat[Sample \textsf{B}]{
\label{tab:corfitBredux}
\centering
\begingroup\footnotesize
\begin{tabular}{rlll}
  \toprule
 & 1 & 2 & 3 \\ 
  \midrule
1. ctrl & 0.646 &  &  \\ 
  2. aware & 0.573 & 0.802 &  \\ 
  3. collect & 0.221 & 0.316 & 0.772 \\ 
   \bottomrule
\multicolumn{4}{c}{\emph{Note:} The diagonal contains the $\sqrt{\mathit{AVE}}$}\\
\end{tabular}
\endgroup
}~\subfloat[Sample \textsf{V}]{%
\label{tab:corfitVredux}
\centering
\begingroup\footnotesize
\begin{tabular}{rlll}
  \toprule
 & 1 & 2 & 3 \\ 
  \midrule
1. ctrl & 0.663 &  &  \\ 
  2. aware & 0.63 & 0.718 &  \\ 
  3. collect & 0.268 & 0.415 & 0.69 \\ 
   \bottomrule
\multicolumn{4}{c}{\emph{Note:} The diagonal contains the $\sqrt{\mathit{AVE}}$}\\
\end{tabular}
\endgroup
}
\end{table*}
}
\newcommand{\htmtBVredux}{
\begin{table*}[p]
\centering\caption{Heterotrait-Monotrait ratios as criterion for discriminant validity on IUIPC-8}
\captionsetup{position=top}
\label{tab:htmtBVredux}
\subfloat[Sample \textsf{B}]{
\label{tab:htmtBredux}
\centering
\begingroup\footnotesize
\begin{tabular}{rlll}
  \toprule
 & 1 & 2 & 3 \\ 
  \midrule
1. ctrl & --- &  &  \\ 
  2. aware & 0.47 & --- &  \\ 
  3. collect & 0.14 & 0.25 & --- \\ 
   \bottomrule
\end{tabular}
\endgroup
}~\subfloat[Sample \textsf{V}]{%
\label{tab:htmtVredux}
\centering
\begingroup\footnotesize
\begin{tabular}{rlll}
  \toprule
 & 1 & 2 & 3 \\ 
  \midrule
1. ctrl & --- &  &  \\ 
  2. aware & 0.55 & --- &  \\ 
  3. collect & 0.23 & 0.34 & --- \\ 
   \bottomrule
\end{tabular}
\endgroup
}
\end{table*}
}



  \title[Validity and Reliability of IUIPC-8]{Towards Valid and Reliable Privacy Concern Scales: The Example of IUIPC-8}
  \titlenote{This is the author's copy of the book chapter of the same name, to appear in the open-access book Nina Gerber, Alina St{\"o}ver and Karola Marky. \textit{Human Factors in Privacy Research}. Springer Verlag, Cham, 2023.} 

  \author{Thomas Gro{\ss}}
  \orcid{0000-0002-7766-2454}
  \affiliation{%
  \institution{Newcastle University}
  \department{School of Computing}
  \city{Newcastle upon Tyne}
  \country{United Kingdom}
  }
 \email{thomas.gross@newcastle.ac.uk}


\maketitle

%
%

Valid and reliable measurement instruments are vital for human factors in privacy research~\cite{john2000measurement}. Validity means that an instrument measures what it purports to measure. Reliability means that the instrument measures this consistently.

In this chapter, we focus on the validity and reliability of privacy concern scales. While there is a range of privacy concern and behavior measurement instruments available~\cite{smith1996information,malhotra2004internet,dinev2004internet,xu2008examining,buchanan2007development,braunstein2011indirect,preibusch2013guide,hong2013internet}, also discussed in studies on the privacy paradox~\cite{kokolakis2017privacy,gerber2018explaining}, we will focus on the scale {Internet Users' Information Privacy Concerns} (IUIPC)~\cite{malhotra2004internet}. IUIPC has roots in the earlier scale {Concerns for Information Privacy} (CFIP)~\cite{smith1996information}, itself a popular scale measuring organizational information privacy concern and validated in independent studies~\cite{stewart2002empirical,harborth2018german}.

IUPIC has been appraised by researchers as part of other studies~\cite{sipior2013empirically,morton2013measuring} and undergone an independent empirical evaluation of the scale itself~\cite{Gross2021IUIPC} and of the applicability of the full nomology in other cultures~\cite{pape2020re}. Even though the scale was originally created in a diligent, evolutionary fashion and founded on a sound underpinning for its content validity, construct validity and internal consistency reliability were not always found up to par for the purpose of human factors in privacy research.

In this chapter, we will discuss a brief form of the Internet Users' Information Privacy Concern scale (IUIPC)~\cite{malhotra2004internet} as a running example. The brief form, called IUIPC-8, only uses eight of the original ten items, and was determined to yield stronger construct validity and internal consistency reliability~\cite{Gross2021IUIPC}.

Our aim for this chapter is not only to present the IUIPC-8 scale itself, but also to shed light on methods for the evaluation of valid and reliable measurement instruments. To that end, we will employ confirmatory factory analysis (CFA) as the tool of choice. We will use CFA to model the ordinal non-normal data of the questionnaire, to confirm the three-dimensionality of IUIPC-8, to establish global and local fit, and finally to estimate construct validity and internal consistency reliability metrics.

We begin this chapter with an overview of information privacy concern and predominant measurement instruments in Section~\ref{sec:priv_concern}. We give a brief introduction of validity and reliability notions in Section~\ref{sec:reqs} and lay the foundations for the use of confirmatory factor analysis as tool to evaluate measurement-instrument properties in Section~\ref{sec:fa}. We discuss the abstract approach used for the evaluation of IUIPC-8 in Section~\ref{sec:approach} and include the empirical results in the validation of the instrument in Section~\ref{sec:results}. Section~\ref{sec:discussion} highlights aspects of the scale properties and considerations for its use in practice in a general discussion.
Definitions used throughout the chapter are summarized in Table~\ref{tab:definitions}.

\begin{table}[b]
\caption{Definitions}
\label{tab:definitions}
\begin{tabular}{p{\textwidth}}
\toprule
\begin{compactitem}[$\blacksquare$]
\item \textbf{\refterm{validity}{Validity}:} Capacity of an instrument to measure what it purports to measure.~\cite{messick1987validity,borsboom2004concept}
\item \textbf{\refterm{reliability}{Reliability}:} Extent to which a variable is consistent in what is being measured.~\cite[p. 123]{hair2019multivariate}
\item \textbf{Construct Validity:} Whether the measure accurately reflects the construct intended to measure.~\cite{messick1987validity,john2000measurement}
\item \textbf{\refterm{factorial_validity}{Factorial Validity}:} Factor composition and dimensionality are sound.
\item \textbf{Confirmatory Factor Analysis:} Factor analysis in a restricted measurement model: each indicator is to depend only on the factors specified.~\cite[p. 191]{kline2015principles}
\item \textbf{Nested Model:} A model that can be derived from another by restricting free parameters.
 \item \textbf{Accept-support test:} A statistical inference, in which the acceptance of the null hypothesis supports the model, e.g., the close-fit test.~\cite[p. 265]{kline2015principles}
 \item \textbf{Reject-support test:} A statistical inference, in which the rejection of the null hypothesis supports the model, e.g., the not-close-fit test.~\cite[p. 265]{kline2015principles}
\item\textbf{Fit Statistic:} A summary measure of the average discrepancy between the sample and model covariances.
\item \textbf{Goodness-of-Fit $\chi^2$:} Measures the exact fit of a model and gives rise to the \refterm{accept-support}{accept-support} exact-fit test against null hypothesis $H_{\chi^2, 0}$.
  \item \textbf{\textsf{RMSEA}:} \refterm{RMSEA}{Root Mean Square Estimate of Approximation}, an absolute badness-of-fit measure estimated as $\hat{\varepsilon}$ with its $90\%$ confidence interval, yielding a range of fit-tests: \refterm{close_fit}{close fit}, \refterm{not_close_fit}{not-close fit}, and \refterm{poor_fit}{poor fit}.~~\cite[pp. 274]{kline2015principles}
  \item \textbf{\refterm{CFI}{Bentler Comparative Fit Index (CFI)}:} An incremental fit index based on the non-centrality measure comparing selected against the null model.
 \item \textbf{\refterm{SRMR}{Standardized Root Mean Square Residual} (SRMR):} A standardized version of the mean absolute covariance residual, for which zero indicates excellent fit.
 \item \textbf{Standardized Factor Loading $\beta$:} $Z$-transformed factor score.
  \item \textbf{Variance Extracted $R^2$:} The factor variance accounted for, computed as squared standardized loading $\beta^2$.
 \item \textbf{\refterm{AVE}{Average Variance Extracted} (AVE):} The average of the squared standardized loadings $\beta^2$ of indicators belonging to the same factor.~\cite[pp. 313]{kline2015principles}
  \item \textbf{\refterm{heterotrait_monotrait_ratio}{Heterotrait-Monotrait (HTMT) Ratio}:} A metric of distriminant validity, the ratio of the avg. correlations of indicators across constructs measuring different phenomena to the avg. correlations of indicators within the same construct~\cite{henseler2015new}.
  \item \textbf{Cronbach's $\alpha$:} Internal consistency based on the average inter-item correlations. 
  \item \textbf{Congeneric Reliability $\omega$:} The amount of general factor saturation (also called \defterm{CR}{composite reliability}~\cite[pp. 313]{kline2015principles} or construct reliability (CR)~\cite[p. 676]{hair2019multivariate} depending on the source).
\end{compactitem}
\\\bottomrule
\end{tabular}
\end{table} 

%

\section{Information Privacy Concern}
\label{sec:priv_concern}

\subsection{What is Information Privacy Concern?}
Malhotra et al.~\cite[p. 337]{malhotra2004internet} ground their definition of \defterm{ipc}{information privacy concern} in Westin's definition of information privacy as a foundation of their understanding of privacy concern: ``the claim of individuals, groups, or institutions to determine for them selves when, how, and to what extent information about them is communicated to others.'' They define information privacy concern as ``an individual's subjective views of fairness within the context of information privacy.'' 

This framing of information privacy concern resonates with the interdisciplinary review of privacy studies by Smith et al.~\cite{smith2011information}. Therein, privacy concern is shown as the central antecedent of related behavior in the privacy macro model.
At the same time, the causal impact of privacy concern on behavior has been under considerable scrutiny. The observed phenomenon, the privacy attitude-behavior dichotomy is commonly called the \defterm{privacy_paradox}{privacy paradox}~\cite{gerber2018explaining}.
Investigating the privacy paradox has been a mainstay topic of the human aspects of privacy community. 
This investigation calls for instruments to measure information privacy concern accurately and reliably, because measurement errors and correlation attenuation of invalid or unreliable privacy concern instruments could confound the research on the privacy paradox.

\subsection{Information Privacy Concern Instruments}
Information privacy concern can be measured by a range of related and distinct instruments~\cite{smith1996information,malhotra2004internet,dinev2004internet,xu2008examining,buchanan2007development,braunstein2011indirect}. As a comprehensive comparison would be beyond the scope of this chapter, we refer to Preibusch's comprehensive guide to measuring privacy concern~\cite{preibusch2013guide} for an overview of the field. We will consider the scales most closely related to IUIPC. Table~\ref{tab:instruments} offers a brief overview of these instruments and their dimensions. While IUIPC is one of the most used internet privacy concern scales, its dimensions also influenced further scales, such as Hong and Thong's Internet Privacy Concern (IPC)~\cite{hong2013internet}. Still, it remains a relatively concise scale.

First point of call is the scale \defterm{CFIP}{Concern for information privacy} (CFIP)~\cite{smith1996information} as a major influence on the development of IUIPC. CFIP consists of four dimensions---\fName{Collection}, \fName{Unauthorized Secondary Use}, \fName{Improper Access} and \fName{Errors}. While both questionnaires share questions, CFIP focuses on individuals' concerns about organizational privacy practices and the organization's responsibilities. CFIP received independent empirical confirmations of its factor structure, by Stewart and Segars~\cite{stewart2002empirical} and by Harborth and Pape~\cite{harborth2018german} on its German translation.

The scale \defterm{IUIPC}{Internet Users' Information Privacy Concern} (IUIPC) was developed by Malhotra et al.~\cite{malhotra2004internet}, by predominately adapting questions of the earlier 15-item scale Concern for Information Privacy (CFIP) by Smith et al.~\cite{smith1996information} and by framing the questionnaire for Internet users as consumers.
IUIPC is measuring their perception of fairness and justice in the context of information privacy and online companies. 
IUIPC-10 was established as a second-order reflective scale of \defterm{dimensions_IUIPC}{information privacy concern}, with the dimensions \fName{Control}, \fName{Awareness}, and \fName{Collection}. The authors considered the ``act of collection, whether it is legal or illegal,'' as the starting point of information privacy concerns. The sub-scale \fName{Control} is founded on the conviction that ``individuals view procedures as fair when they are vested with control of the procedures.'' The authors considered being ``informed about data collection and other issues'' as central concept of the sub-scale \fName{Awareness}.

Initial appraisals of IUIPC-10~\cite{sipior2013empirically,morton2013measuring} yielded concerns for the validity and reliability of the scale largely tied to two items on awareness and control. These validity and reliability problems were confirmed in an independent empirical evaluation of the scale~\cite{Gross2021IUIPC}. Pape et al.~\cite{pape2020re} independently evaluated the full nomology of IUIPC-10 in Japan.

\defterm{IPC}{Internet Privacy Concerns} (IPC)~\cite{dinev2004internet} considered internet privacy concerns with antecedents of perceived vulnerability and control, antecedents familiar from the Protection Motivation Theory (PMT). IPC differs from IUIPC in its focus on misuse rather than just collection of information and of concerns of surveillance. 
In terms of the core scale of privacy concern, Dinev and Hart identified two factors 
\begin{inparaenum}[(i)] 
\item \fName{Abuse} (concern about misuse of information submitted on the Internet) and
\item \fName{Finding} (concern about being observed and specific private information being found out).
\end{inparaenum}
It considered the two antecedents \fName{Control} and \fName{Vulnerability}.
The IPC scale was subsequently expanded on and integrated with other scales by Hong and Thong~\cite{hong2013internet} and further investigated with respect to four driver and inhibitor dimensions by Hong et al.~\cite{hong2021drivers}. Herein, Hong and Thong reformulated questions to more consistently express concern.

Buchanan et al.'s \defterm{OPC}{Online Privacy Concern and Protection for Use on the Internet} (OPC)~\cite{buchanan2007development} measure considered three sub-scales---\fName{General Caution}, \fName{Technical Protection} (both on behaviors), and \fName{Privacy Attitude}. Compared to IUIPC, OPC sports a strong focus on item stems eliciting being concerned and on measures regarding a range of concrete privacy risks.

\begin{table}[tbp]
\caption{Overview of selected privacy concern instruments}
\label{tab:instruments}
\begin{tabular}{p{5cm}clcc}
\toprule
Instrument & Year & Dimensions & Development & Appraisals\\
\midrule
\multirow{4}{*}{\parbox{5cm}{Concern for information privacy  (CFIP)}} & \multirow{4}{*}{1996} & \fName{Collection\textsubscript{a}} & \multirow{4}{*}{\cite{smith1996information}} & \multirow{4}{*}{\cite{stewart2002empirical,harborth2018german}} \\
 && \fName{Unauthorized Secondary Use\textsubscript{b}}\\
 && \fName{Improper Access\textsubscript{c}}\\
 && \fName{Errors\textsubscript{d}}\\
 \midrule
\multirow{3}{*}{\parbox{5cm}{Internet Users' Information Privacy Concern (IUIPC)}} & \multirow{3}{*}{2004} & \fName{Control\textsubscript{f}} & \multirow{3}{*}{\cite{malhotra2004internet}} & \multirow{3}{*}{\cite{sipior2013empirically,morton2013measuring,pape2020re,Gross2021IUIPC}}\\
&&\fName{Awareness\textsubscript{e}}\\
&&\fName{Collection\textsubscript{a}}\\
\midrule
\multirow{4}{*}{\parbox{5cm}{Internet Privacy Concerns (IPC)}}  & \multirow{4}{*}{2004} &  \fName{Control} & \multirow{4}{*}{\cite{dinev2004internet}} & \\
 && \fName{Vulnerability}\\
 && \fName{Abuse}\\
 && \fName{Finding}\\
\midrule
\multirow{6}{*}{\parbox{5cm}{Internet Privacy Concerns (IPC)}} & \multirow{6}{*}{2013} & \fName{Collection\textsubscript{a}} & \multirow{6}{*}{\cite{hong2013internet}} & \multirow{6}{*}{\cite{terlizzi2019replication,hong2021drivers}} \\
&& \fName{Secondary Usage\textsubscript{b'}} \\
&& \fName{Errors\textsubscript{d'}} \\
&& \fName{Improper Access\textsubscript{c'}} \\
&& \fName{Control\textsubscript{f}} \\
&& \fName{Awareness\textsubscript{e}} \\
\midrule
\multirow{3}{*}{\parbox{5cm}{Online Privacy Concern and Protection for Use on the Internet (OPC)}} & \multirow{3}{*}{2007} & \fName{General Caution} & \multirow{3}{*}{\cite{buchanan2007development}} \\
 && \fName{Technical Protection} \\
 && \fName{Privacy Attitude}\\
\bottomrule
\end{tabular}
\\\emph{Note:} Dimensions with the same subscript bear relations to another. A prime indicates that the items of the dimension were considerably reformulated.
\end{table} 

\section{Validity and Reliability}
\label{sec:reqs}

When evaluating a privacy concern instrument such as IUIPC-8, the dual vital questions for research in human factors of privacy and the privacy paradox are:
\begin{inparaenum}[(i)]
  \item Are we measuring the hidden latent construct \fName{Privacy Concern} accurately? (validity)
  \item Are we measuring privacy concern consistently and with an adequate signal-to-noise ratio? (reliability)
\end{inparaenum}
Without sufficient reliability, a measurement instrument cannot be valid~\cite{john2000measurement}.

\defterm{validity}{Validity} refers to whether an instrument measures what it purports to measure. Messick offered an early well-regarded definition of validity as the ``integrated evaluative judgment of the degree to which empirical evidence and theoretical rationales support the adequacy and appropriateness of inferences and actions based on test scores''~\cite{messick1987validity}. Validity is inferred---judged in degrees---not measured.
In this chapter, we put our attention on the validation procedure and underlying evidence for validity and reliability.
In that, we largely take content validity of IUIPC for granted. \defterm{content_validity}{Content validity} refers to the relevance and representativeness of the content of the instrument, typically assessed by expert judgment.

\subsection{Construct Validity}
\label{sec:construct_validity}

Messick~\cite{messick1980test} defines \defterm{construct_validity}{construct validity}~\cite{cronbach1955construct}, the interpretive meaningfulness, as the extent to which an instrument accurately represents a construct. This definition has also been used in more recent papers on measurement~\cite{john2000measurement} as a primary kind of validity. Construct validity is typically established by the evaluation of the instrument through multiple lenses, where we will go into factorial, convergent and discriminant validity.

\paragraph*{Factorial Validity}
First, we seek evidence of \defterm{factorial_validity}{factorial validity}, that is, evidence that that factor composition and dimensionality are sound. While IUIPC is a \defterm{multidimensional}{multidimensional} scale with three correlated designated dimensions, we require \defterm{unidimensionality}{unidimensionality} of each sub-scale, a requirement discussed at length by Gerbing and Anderson~\cite{gerbing1988updated}. 

Unidimensional measurement models for sub-scales correspond to expecting \defterm{congeneric}{congeneric} measures, that is, the scores on an item are the expression of a true score weighted by the item's loading plus some measurement error, where in the congeneric case neither the loadings nor error variances across items are required to be equal.
This property entails that the items of each sub-scale must be conceptually homogeneous. 

We find empirical evidence for factorial validity of a scale's measurement model in the closeness of fit to the sample's covariance structure. Specifically, we gain supporting evidence by passing fit hypotheses of a confirmatory factor analysis for the designated factor structure~\cite{anderson1987assessment,gerbing1988updated,kline2015principles}, where we prioritize fit metrics and hypotheses based on the \refterm{RMSEA}{RMSEA} included in Table~\ref{tab:hypotheses}.

\paragraph*{Convergent and Discriminant Validity}
\defterm{convergent_validity}{Convergent validity}~\cite[pp. 675]{hair2019multivariate} (convergent coherence) on an item-construct level means that items belonging together, that is, to the same construct, should be observed as related to each other.
Similarly, \defterm{discriminant_validity}{discriminant validity}~\cite[pp. 676]{hair2019multivariate} (discriminant distinctiveness) means that items not belonging together, that is, not belonging to the same construct, should be observed as not related to each other. On a sub-scale level, we expect factors of the same higher-order construct to relate to each other and, on hierarchical factor level, we expect all 1\textsuperscript{st}-order factors to load strongly on the 2\textsuperscript{nd}-order factor.

In first instance, a poor local fit and tell-tale residual patterns yield disconfirming evidence for convergent and discriminant validity. We can further inspect inter-item correlation matrices: we expect items belonging to the same sub-scale to be highly correlated and, thereby, to converge on the same construct.  Correlation to items of other sub-scales should be low, especially lower than the in-construct correlations~\cite[pp. 196]{kline2015principles}. 

These principles give rise to criteria based on the Average Variance Extracted (AVE), the Fornell-Larcker criterion~\cite{fornell1981evaluating} and the \refterm{heterotrait_monotrait_ratio}{Heterotrait-Monotrait Ratio} (HTMT)~\cite{henseler2015new,ab2017discriminant}. We summarize these terms in Table~\ref{tab:definitions}.

\subsection{Reliability}
\label{sec:reliability}
\defterm{reliability}{Reliability} is the extent to which a variable is consistent in what is being measured~\cite[p. 123]{hair2019multivariate}. It can further be understood as the capacity of ``separating signal from noise''~\cite[p. 709]{RevCon2018,john2000measurement}, quantified by the ratio of true score to observed score variance.~\cite[pp. 90]{kline2015principles}
We evaluate \defterm{internal_consistency}{internal consistency} as a means to estimate reliability from a single test application.
Internal consistency entails that items that purport to measure the same construct produce similar scores~\cite[p. 91]{kline2015principles}.
We will use the internal consistency measures Cronbach's~$\alpha$, congeneric reliability~$\omega$, and AVE, defined in Table~\ref{tab:definitions}. While Cronbach's~$\alpha$ is well-known in the community, Average Variance Extracted (AVE) offers a simple intuitive measure and congeneric reliability provides a robust approach.

Thresholds for reliability estimates like Cronbach's $\alpha$ or Composite Reliability $\omega$ are debated in the field, where many recommendations are based on Nunnally's original treatment of the subject, but equally often misstated~\cite{lance2006sources}. The often quoted $\alpha \geq .70$ was described by Nunnally only to ``save time and energy,'' whereas a greater threshold of $.80$ was endorsed for basic research~\cite{lance2006sources}. 

When we designate \textit{a priori} thresholds as criteria for internal consistency reliability, this approach needs to be put into a broader context. As for validity, reliability is judged in degrees. John and Benet-Mart{\'i}nez~\cite{john2000measurement} discuss the arbitrariness of one-size-fits-all fixed reliability thresholds. Internal consistency reliability needs to be considered in relation to inter-item correlations and the length of a scale and, further, how these aspects fit the nature of the construct in question.
Ultimately, the choice of thresholds gives rise to a bandwidth-fidelity trade-off~\cite{john2000measurement}. Whether we call an instrument ``reliable enough'' depends on the proportion of error variance we are willing to tolerate and on the attenuation of the correlation to other variables as a consequence of that.  

\section{Factor Analysis as Tool to Establish Measurement Instruments}
\label{sec:fa}
Factor analysis is a powerful tool for evaluating the \refterm{construct_validity}{construct validity} and \refterm{reliability}{reliability} of privacy concern instruments. Thereby, it constitutes validation procedure for the measurement instruments~\cite{john2000measurement}.
\defterm{FA}{Factor analysis} refers to a set of statistical methods that are meant to determine the number and nature of \defterm{LV}{latent variables} (LVs) or \defterm{factor}{factors} that account for the variation and covariation among a set of observed measures commonly referred to as \defterm{IV}{indicators}~\cite{brown2015confirmatory}.

\begin{figure}
\centering
\includegraphics[keepaspectratio,width=.5\textwidth]{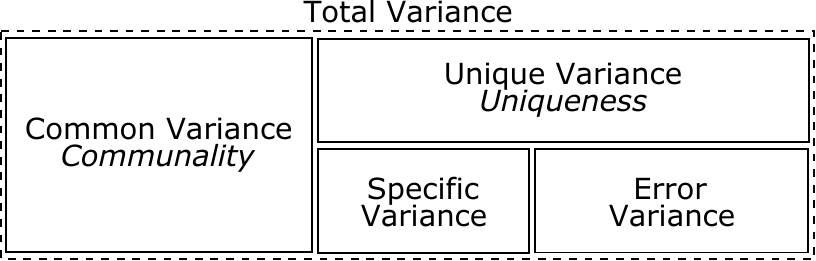}
\caption{Variance attribution in the Common Factor Model (CFM)}
\label{fig:common_factors}
\Description[Overview of the Common Factor Model showing how variance is attributes to different sources.]{The image shows the attribution of variance in the Common Factor Model. The total variance of indicator variables is attributed on the one hand to common variance of the latent factor, also called communality. There is further variance unique to an indicator, called uniqueness. This variance is either specific variance from the item or error variance.}
\end{figure}

\defterm{CFA}{Confirmatory factor analysis} (CFA) is a factor analysis in a restricted measurement model, that is, in which each indicator is to depend only on the factors specified.~\cite[pp. 191]{kline2015principles}. CFA is commonly used to evaluate psychometric instruments. It is based on the \defterm{CFM}{common factor model (CFM)}, which holds that each indicator variable contributes to the variance of one or more common factors and one unique factor. Thereby, \defterm{common variance}{common variance} of related observed measures is attributed to the corresponding latent factor, and \defterm{unique variance}{unique variance} (uniqueness) seen either as variance associated with the item or as error variance. We call the proportion of variance associated with factors \defterm{communality}{communality}, the proportion of variance not associated with factors \defterm{uniqueness}{uniqueness}. We depict the common factor model in Figure~\ref{fig:common_factors}. 

IUIPC is based on a \defterm{reflective}{reflective measurement}, that is, the observed measure of an indicator variable is seen as \emph{caused} by some latent factor. 
Indicators are thereby \defterm{endogenous}{endogenous} variables, latent variables \defterm{exogenous}{exogenous} variables.
Reflective measurement requires that all items of the sub-scale are interchangeable~\cite[pp. 196]{kline2015principles}.
In this chapter, we focus on \defterm{CB-CFA}{covariance-based confirmatory factor analysis} (CB-CFA). Therein, the statistical tools aim at estimating coefficients for parameters of the measurement model that best fit the covariance matrix of the observed data.
The difference between an observed covariance of the sample and an implied covariance of the model is called a \defterm{residual}{residual}.

\subsection{Estimation Methods for Ordinal Non-Normal Data}
\label{sec:est_assumptions}
The purpose of a \refterm{FA}{factor analysis} is to estimate free parameters of the model (such as loadings or error variance), which is facilitated by \defterm{estimator}{estimators}.
The choice of estimator matters, because each comes with different strengths and weaknesses, requirements and assumptions that need to be fulfilled for the validity of their use. 

While \defterm{ML}{maximum likelihood} (ML) estimation is the model commonly used estimation method for CFA, it is based on \defterm{ml_assumptions}{assumptions}~\cite[pp. 71]{kline2015principles} that are not satisfied by IUIPC:
\begin{inparaenum}[(i)]
  \item a \defterm{continuous}{continuous measurement level},
  \item \defterm{normal_distribution}{multi-variate normal distribution} (entailing the absence of extreme \defterm{skewness}{skewness})~\cite[pp. 74]{kline2015principles}.
\end{inparaenum}
The distribution requirements are placed on the \refterm{endogenous}{endogenous} variables: the indicators.

First, the Likert items are used in IUIPC are \defterm{ordinal}{ordinal}~\cite[p. 11]{hair2019multivariate}, that is, ordered categories in which the distance between categories is not constant. Lei and Wu~\cite{LeiWu2012} held based on a number of empirical studies that the fit indices of approximately normal \refterm{ordinal}{ordinal} variables with at least five categories are not greatly misleading. However, when ordinal and non-normal is treated as continuous and normal, the fit is underestimated and there is a more pronounced negative bias in estimates and standard errors. While Bovaird and Kozoil~\cite{bovaird2012measurement} acknowledge robustness of the ML estimator with \refterm{normal_distribution}{normally distributed} \refterm{ordinal}{ordinal} data, they stress that increasingly skewed and kurtotic ordinal data inflate the Type I error rate, hence require another approach~\cite[pp. 323]{kline2015principles}. In the same vein, Kline~\cite[p. 122]{kline2012assumptions} holds the normality assumption for \refterm{endogenous}{endogenous} variables---the indicators---to be critical.


\subsection{Comparing Nested Models} \defterm{nested_model}{Nested models}~\cite[p. 280]{kline2015principles} are models that can be derived from each other by restricting free parameters. They can be compared with a \defterm{LRT}{Likelihood Ratio $\chi^2$ Difference Test} (LRT)~\cite[p. 270]{kline2015principles}. This technique comes into play when we compare multiple models that are based on the same indicator variables, e.g., to establish which factor structure most suits the covariance matrix. We use this technique in comparing one-factor solutions with solutions with multiple factors.

\subsection{Global and Local Fit}
\label{sec:fit.tests}

The \defterm{fit}{closeness of fit} of a factor model to an observed sample is evaluated globally with fit indices as well as locally by inspecting the residuals.
We shall focus on the ones Kline~\cite[p. 269]{kline2015principles} required as minimal reporting.

\paragraph*{Statistical Inference} 
The $\chi^2$ and \textsf{RMSEA} indices offer us \defterm{fit_tests}{statistical inferences of global fit}. Such tests can either be \defterm{accept-support}{accept-support}, that is, accepting the null hypothesis supports the selected model, or \defterm{reject-support}{reject-support}, that is, rejecting the null hypothesis supports the selected model. We present them in Table~\ref{tab:hypotheses}.

\paragraph*{Local Fit}
Even with excellent global fit indices, the inspection of the local fit---evidenced by the residuals---must not be neglected.  Kline~\cite[p. 269]{kline2015principles} emphasises ``Any report of the results without information about the residuals is incomplete.''
Large absolute \refterm{residual}{residuals} indicate covariations that the model does not approximate well and that may, thereby, lead to spurious results. 

\section{Approach}
\label{sec:approach}

In this section, we are weaving a general approach for creating a valid and reliable measurement instrument with specific design decisions taken for the brief information privacy concerns scale IUIPC-8~\cite{Gross2021IUIPC}. General approaches for systematic constructions of measurements~\cite{john2000measurement}, measurement models for survey research~\cite{Bohrnstedt2010}, or their reliability and validity~\cite{Alwin2010} are well documented in literature.
Here, we introduce specific considerations for IUIPC-8.

The following aspects inform this evaluation:
\begin{compactitem}[$\blacksquare$]
  \item The scale IUIPC-8 is derived from the long-standing scale IUIPC-10. Hence, a comparison of both scales is in order.
  \item We will conduct confirmatory factor analyses to establish the dimensionality and construct validity of the scale.
  \item The IUIPC data will be from ordinal Likert items, with a skewed non-normal distribution.
  \item We will need sufficient sample sizes for the statistical power on RMSEA-based statistical inferences.
  \item We aim at a scale that yields low attenuation of the correlations of its latent variable in the relation to other variables, requiring good internal consistency reliability.
\end{compactitem}

\subsection{Analysis Methodology}

\begin{figure}
\centering
\includegraphics[keepaspectratio, width=\columnwidth]{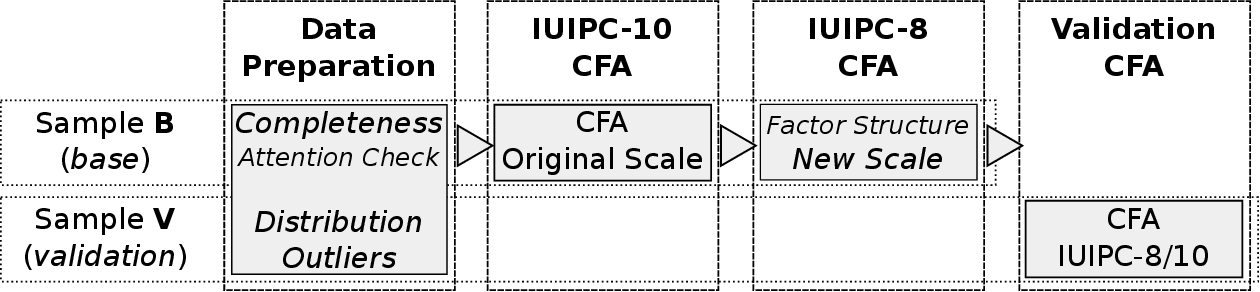}
\caption{Which steps were taken on what sample (adapted from~\cite{Gross2021IUIPC})}
\label{fig:process}
\Description[The stages of the study showing which steps are taken on which sample.]{The study progressed over four stages, in which steps were taken on base sample B and validation sample V. In the data preparation stage, the sample was refined checking for completeness and successful attention checks as well as fulfilling distribution and outlier assumptions. This was done on both samples. In the stage IUIPC-10 CFA, the 10-item IUIPC scale was analysed with Confirmatory Factor Analysis (CFA) on base sample B. Then, in stage IUiPC-8 CFA, a new factor structure was proposed that eliminated two items showing weaknesses. In this stage another Confirmatory Factor Analysis checked the new model. in the final stage, Validation CFA, the models were analyzed on the independent validation sample V.}
\end{figure}

Our analysis of IUIPC-10 and the brief variant IUIPC-8 will be supported by confirmatory factor analyses on two independent samples, one used for specification and refinement, the other used for validation. The factor analyses yield the evidence for uni-dimensionality of the sub-scales and the overall dimensionality of the instrument. While the creation of a new measurement instrument would often start with an exploratory factor analysis on a candidate item pool and another independent sample, here we shall focus only on the confirmatory factor analyses setting the 8-item and 10-item variants apart.
The corresponding analysis process is depicted in Figure~\ref{fig:process}.

Because IUIPC yields ordinal, non-normal data, the distribution of the data asks for careful analysis as part of the data preparation. 
The assumptions of a standard maximum likelihood estimation will be violated, by which we are preparing for a robust diagonally weighted least square (DWLS) estimation method as tool of choice. 
The specific method employed is called \defterm{wlsmv}{WLSMV}, a robust diagonally weighted least square (DWLS) estimation with robust standard errors and mean- and variance-adjusted test statistics using a scale-shift.
The choice of estimation method will also impact the sample size we need to obtain: Apart from cases of small samples ($N < 200$), WLSMV was found to be less biased and more accurate than robust ML estimation (MLR)~\cite{ChengHsien2016}. For smaller sample sizes, we would recommend MLR.

\subsection{Sample}
The quality of the sampling method, that is, how participants are drawn from a population have a considerable impact on the sampling and non-sampling biases introduced early in a study. In an ideal case, the target and survey population are clearly specified and the sampling frame explicitly defined.~\cite{henry1990practical,williams1978sampler}
In terms of sampling method random sampling, possibly stratified to be representative of the population, carries the least bias.

The sample size is determined based on three requirements:
\begin{inparaenum}[(i)] 
  \item the size needed to reach a diversity approximately representative of the UK population ($N > 300$),
  \item the minimum sample size to make DWLS estimation viable ($N > 200$), and
  \item the sample size required to reach adequate statistical power. 
\end{inparaenum}
For confirmatory factor analyses considered in this chapter the key statistical inferences are on the $\chi^2$ significance test for the exact fit and the RMSEA-based significance tests for approximate fit. Hence, we determined the sample size for an RMSEA close-fit test in an \textit{a priori} power analysis~\cite{wolf2013sample,maccallum1996power,kyriazos2018applied}. We used the \textsf{R} package \textsf{SEMPower} and an IUIPC-10 model with $n_{\const{par}} = 23$ free parameters and $\vari{df} = 32$ degrees of freedom as benchmark. To reach $1-\beta = 80\%$ statistical power in this constellation, we would need a target sample size of $N_{1-\beta =.80} = 317$.

For the analysis of IUIPC-8,  we employed two independent samples, \textsf{B} and \textsf{V}.
Base sample~\textsf{B} and validation sample~\textsf{V} were designated with a sample size of $420$ each, allowing for some sample size and power loss in the sample refinement and analysis. 

The samples used here were used for an earlier study~\cite{Gross2021IUIPC} establishing IUIPC-8, hence serve for illustration and not as an independent validation of the questionnaire.
The samples were recruited on Prolific Academic~\cite{palan2018prolific} to be representative of the UK census by age, gender and ethnicity. The sampling frame was Prolific users who were registered on the platform as residents of the UK, consisting of $48,454$ users at sampling time (August 2019).
The sampling process was as follows:
\begin{inparaenum}[1.]
  \item Prolific established sample sizes per demographic strata of the intended population,
  \item it presented our studies to the registered users with matching demographics,
  \item the users could choose themselves whether they would participate or not.
\end{inparaenum}
We enforced sample independence by uniqueness of the participants' Prolific ID.

We note that the sampling method is not random, it is a crowdsourcing sample with demographics screening~\cite{landers2015inconvenient}, yet Prolific has been found to obtain samples from a diverse population and with high data quality and reliability~\cite{peer2017beyond}.

\subsection{Validity and Reliability Criteria}
\label{sec:criteria}
The first consideration for construct validity is in the factorial validity of the model, where we compare multiple possible factor structures to confirm the dimensionality. Based on the overall selected model structure, we then turn to an analysis of the global and local fit in a comparison between IUIPC-10 and IUIPC-8 on samples \textsf{B} and \textsf{V}. For the fit of the models, we consider the RMSEA-based fit hypotheses shown in Table~\ref{tab:hypotheses} as important part of our investigation. Here we are interested in getting support from the close-fit hypothesis, being aware that the CFAs will not have enough statistical power to offer a tight enough confidence interval on the RMSEA-estimate to reject the not-close fit hypothesis.

\begin{table}
\centering
\caption{Exact and approximate fit hypotheses}
\label{tab:hypotheses}
\begin{tabular}{llcc}
\toprule
Type & Null Hypothesis & Classification & \textsf{RMSEA} 90\% CI  \\
\midrule
Exact fit & $H_0 : \varepsilon_0 = 0$ & accept-support & $\hat{\varepsilon}_{\mathsf{L}} = 0$ \\
Close fit & $H_0 : \varepsilon_0 \leq .05$ & accept-support & $\hat{\varepsilon}_{\mathsf{L}} \leq .05$\\
Not-close fit & $H_0 : \varepsilon_0 \geq .05$ & reject-support & $\hat{\varepsilon}_{\mathsf{U}} < .05$\\
Poor fit & $H_0 : \varepsilon_0 \geq .10$ & reject-support & $\hat{\varepsilon}_{\mathsf{U}} < .10$ \\
\bottomrule
\end{tabular}
\\\emph{Note:} $\varepsilon_0$ = RMSEA under the null hypothesis;\\$\hat{\varepsilon}$ = point estimate of RMSEA; $[\hat{\varepsilon}_{\mathsf{L}}, \hat{\varepsilon}_{\mathsf{U}}]$ = $90\%$ Confidence Interval on $\hat{\varepsilon}$.
\end{table}

For convergent and discriminant validity we turn to empirical criteria, especially relying on the Average Variance Extracted (AVE) and Heterotrait-Monotrait Ratio (HTMT) in Table~\ref{tab:definitions}.
We gain empirical evidence in favor of convergent validity~\cite[pp. 675]{hair2019multivariate}
\begin{inparaenum}[(i)]
  \item if the variance extracted by an item $R^2 > .50$ entailing that the standardized factor loading are significant and $\beta > .70$, and
   \item if the \refterm{internal_consistency}{internal consistency} (defined in Section~\ref{sec:reliability}) is sufficient ($\vari{AVE} > .50$, $\omega > \vari{AVE}$, and $\omega > .70$).
\end{inparaenum}
The analysis yields empirical evidence of discriminant validity~\cite[pp. 676]{hair2019multivariate}
\begin{inparaenum}[(i)]
   \item if the square root of \vari{AVE} of a latent variable is greater than the max correlation with any other latent variable (Fornell-Larcker criterion~\cite{fornell1981evaluating}),
   \item if the \refterm{heterotrait_monotrait_ratio}{Heterotrait-Monotrait Ratio} (HTMT) is less than $.85$~\cite{henseler2015new,ab2017discriminant}.
\end{inparaenum}

While that would be beneficial for privacy research as well, we shall adopt reliability metrics $\alpha, \omega \geq .70$ as suggested by Hair et al.~\cite[p. 676]{hair2019multivariate}. 


\section{The Validation of IUIPC-8}
\label{sec:results}
In this section, we are examining the a model of IUIPC-8 in a Diagonally Weighted Least Square (DWLS) CFA estimation with robust standard errors and a scale-shifted mean- and variance-adjusted test statistic (WLSMV).
We begin our inquiry with the characteristics of the underlying sample (Section~\ref{sec:sample}) and distribution (Section~\ref{sec:desc}), considering a base Sample~\textsf{B} and an independent validation Sample~\textsf{V}.

\subsection{Sample}
\label{sec:sample}

The demographics of both samples \textsf{B} and \textsf{V} are included in Table~\ref{tab:demoCombined}. While these samples were meant to be drawn to be UK representative, we observe an under-representation of elderly participants compared to the UK census age distribution. Still, the sample offers us sufficient diversity for the evaluation of the scale.

%
%
\newcommand{\demoCombinedAll}{
\begin{table}[tb]
\centering\caption{Demographics of Samples \textsf{B} and \textsf{V}}
\captionsetup{position=top}
\label{tab:demoCombinedAll}
\subfloat[Sample \textsf{B}]{
\label{tab:demoCombinedB}
\centering
\begin{tabular}{ll}
\toprule
  & Overall\\
\midrule
$N_{\mathsf{B}}$ & 379\\
Gender (\%) & \\
Female & 197 (52.0)\\
Male & 179 (47.2)\\
Rather not say & 3 ( 0.8)\\
\addlinespace
Age (\%) & \\
18-24 & 41 (10.9)\\
25-34 & 72 (19.0)\\
35-44 & 84 (22.2)\\
\addlinespace
45-54 & 57 (15.0)\\
55-64 & 97 (25.6)\\
65+ & 28 ( 7.4)\\
Mothertongue (\%) & \\
English & 374 (98.7)\\
\addlinespace
Greek & 1 ( 0.3)\\
Portuguese & 1 ( 0.3)\\
Tagalog & 1 ( 0.3)\\
Other & 2 ( 0.5)\\
\bottomrule
\end{tabular}
}~\subfloat[Sample \textsf{V}]{%
\label{tab:demoCombinedV}
\centering
\begin{tabular}{ll}
\toprule
  & Overall\\
\midrule
$N_{\mathsf{V}}$ & 433\\
Gender (\%) & \\
Female & 217 (50.1)\\
Male & 212 (49.0)\\
Rather not say & 4 ( 0.9)\\
\addlinespace
Age (\%) & \\
18-24 & 92 (21.2)\\
25-34 & 143 (33.0)\\
35-44 & 83 (19.2)\\
\addlinespace
45-54 & 58 (13.4)\\
55-64 & 44 (10.2)\\
65+ & 13 ( 3.0)\\
Mothertongue (\%) & \\
English & 422 (97.5)\\
Arabic & 1 ( 0.2)\\
\addlinespace
Greek & 2 ( 0.5)\\
Hindi & 1 ( 0.2)\\
Italian & 2 ( 0.5)\\
Polish & 1 ( 0.2)\\
Russian & 1 ( 0.2)\\
\addlinespace
Spanish & 1 ( 0.2)\\
Urdu & 1 ( 0.2)\\
Other & 1 ( 0.2)\\
\bottomrule
\end{tabular}
}
\end{table}
}

\newcommand{\demoCombined}{
\begin{table}[tb]
\centering\caption{Demographics~\cite{Gross2021IUIPC}}
\captionsetup{position=top}
\label{tab:demoCombined}
\subfloat[Sample \textsf{B}]{
\label{tab:demoCombinedB}
\centering
\begin{tabular}{ll}
\toprule
  & Overall\\
\midrule
$N_{\mathsf{B}}$ & 379\\
Gender (\%) & \\
Female & 197 (52.0)\\
Male & 179 (47.2)\\
Rather not say & 3 ( 0.8)\\
\addlinespace
Age (\%) & \\
18-24 & 41 (10.9)\\
25-34 & 72 (19.0)\\
35-44 & 84 (22.2)\\
\addlinespace
45-54 & 57 (15.0)\\
55-64 & 97 (25.6)\\
65+ & 28 ( 7.4)\\
\bottomrule
\end{tabular}
}~\subfloat[Sample \textsf{V}]{%
\label{tab:demoCombinedV}
\centering
\begin{tabular}{ll}
\toprule
  & Overall\\
\midrule
$N_{\mathsf{V}}$ & 433\\
Gender (\%) & \\
Female & 217 (50.1)\\
Male & 212 (49.0)\\
Rather not say & 4 ( 0.9)\\
\addlinespace
Age (\%) & \\
18-24 & 92 (21.2)\\
25-34 & 143 (33.0)\\
35-44 & 83 (19.2)\\
\addlinespace
45-54 & 58 (13.4)\\
55-64 & 44 (10.2)\\
65+ & 13 ( 3.0)\\
\bottomrule
\end{tabular}
}
\end{table}
}

\newcommand{\demoCombinedWithA}{
\begin{table*}[tb]
\centering\caption{Demographics}
\captionsetup{position=top}
\label{tab:demoCombinedWithA}
\subfloat[Sample \textsf{A}]{
\label{tab:demoCombinedA}
\centering
\begin{tabular}{ll}
\toprule
   & Overall\\
   \midrule
   $N_{\mathsf{A}}$ & 205\\
Gender (\%) & \\
Female & 80 (39.0) \\
Male & 125 (61.0)\\
Rather not say & 0 ( 0.0)\\
\addlinespace
Age (\%) & \\
18-24 & 109 (53.2)\\
25-34 & 71 (34.6)\\
35-44 & 18 ( 8.8)\\
\addlinespace
45-54 & 4 ( 2.0)\\
55-64 & 3 ( 1.5)\\
65+ &    0 ( 0.0)\\
\bottomrule
\end{tabular}
}
\subfloat[Sample \textsf{B}]{
\centering
\label{tab:demoCombinedB}
\centering
\begin{tabular}{ll}
\toprule
  & Overall\\
\midrule
$N_{\mathsf{B}}$ & 379\\
Gender (\%) & \\
Female & 197 (52.0)\\
Male & 179 (47.2)\\
Rather not say & 3 ( 0.8)\\
\addlinespace
Age (\%) & \\
18-24 & 41 (10.9)\\
25-34 & 72 (19.0)\\
35-44 & 84 (22.2)\\
\addlinespace
45-54 & 57 (15.0)\\
55-64 & 97 (25.6)\\
65+ & 28 ( 7.4)\\
\bottomrule
\end{tabular}
}~\subfloat[Sample \textsf{V}]{%
\label{tab:demoCombinedV}
\centering
\begin{tabular}{ll}
\toprule
  & Overall\\
\midrule
$N_{\mathsf{V}}$ & 433\\
Gender (\%) & \\
Female & 217 (50.1)\\
Male & 212 (49.0)\\
Rather not say & 4 ( 0.9)\\
\addlinespace
Age (\%) & \\
18-24 & 92 (21.2)\\
25-34 & 143 (33.0)\\
35-44 & 83 (19.2)\\
\addlinespace
45-54 & 58 (13.4)\\
55-64 & 44 (10.2)\\
65+ & 13 ( 3.0)\\
\bottomrule
\end{tabular}
}
\\\footnotesize{\emph{Note:} Samples \textsf{B} and \textsf{V} were drawn to be representative of the UK census by age and gender; Sample \textsf{A} was not.}
\end{table*}
}

\newcommand{\demoCombinedExp}{
\begin{table*}[tb]
\centering\caption{Demographics}
\captionsetup{position=top}
\label{tab:demoCombinedExp}
\subfloat[Sample \textsf{B}]{
\centering
\label{tab:demoCombinedB}
\centering
\begin{tabular}{ll}
\toprule
  & Overall\\
\midrule
$N_{\mathsf{B}}$ & 379\\
Gender (\%) & \\
Female & 197 (52.0)\\
Male & 179 (47.2)\\
Rather not say & 3 ( 0.8)\\
\addlinespace
Age (\%) & \\
18-24 & 41 (10.9)\\
25-34 & 72 (19.0)\\
35-44 & 84 (22.2)\\
\addlinespace
45-54 & 57 (15.0)\\
55-64 & 97 (25.6)\\
65+ & 28 ( 7.4)\\
\bottomrule
\end{tabular}
}~\subfloat[Sample \textsf{V}]{%
\label{tab:demoCombinedV}
\centering
\begin{tabular}{ll}
\toprule
  & Overall\\
\midrule
$N_{\mathsf{V}}$ & 433\\
Gender (\%) & \\
Female & 217 (50.1)\\
Male & 212 (49.0)\\
Rather not say & 4 ( 0.9)\\
\addlinespace
Age (\%) & \\
18-24 & 92 (21.2)\\
25-34 & 143 (33.0)\\
35-44 & 83 (19.2)\\
\addlinespace
45-54 & 58 (13.4)\\
55-64 & 44 (10.2)\\
65+ & 13 ( 3.0)\\
\bottomrule
\end{tabular}
}
\\\footnotesize{\emph{Note:} Samples were drawn to be representative of the UK census by age and gender}
\end{table*}
}
\demoCombined

The two samples have undergone a sample refinement in stages, which Table~\ref{tab:sample} accounts for. 
The refinement included
\begin{inparaenum}[(i)]
\item removing incomplete cases without replacement,
\item removing duplicates across samples by the participants' Prolific ID, to guarantee independence, and
\item removing cases in which participants failed more than one attention check ($\mathsf{FailedAC} > 1$).
\end{inparaenum}
The named attention checks were Instructional Manipulation Checks~\cite{oppenheimer2009instructional} distributed over the wider questionnaire.

Overall, of the $N_{\mathsf{C}} = 848$ complete cases, only $4.2 \%$ were removed due to duplicates or failed attention checks. After this refinement a small number of multi-variate outliers was removed.

\begin{table}
\centering
\caption{Sample Refinement~\cite{Gross2021IUIPC}}
\label{tab:sample}
\begin{adjustbox}{max width=\columnwidth}
\begin{tabular}{lc@{ }cc@{ }c}
\toprule
\multirow{2}{*}{Phase} &  \multicolumn{2}{c}{\textsf{B}} & \multicolumn{2}{c}{\textsf{V}}\\
\cmidrule(lr){2-3} \cmidrule(lr){4-5} 
  & Excl. & Size & Excl. & Size\\ 
\midrule
Starting Sample & & $473$ & & $467$\\
Incomplete & $58$ & $415$ 
   & $34$ & $433$\\
Duplicate & $25$ & $390$ 
   & $0$ & $433$\\
\textsf{FailedAC}$ > 1$ & $11$ & $379$ 
   & $0$ & $433$\\
MV Outlier & $9$ & $370$ 
   & $14$ & $419$\\
\midrule
Final Sample & \multicolumn{2}{r}{$N_{\mathsf{B}}^\prime = 370$} &  
               \multicolumn{2}{r}{$N_{\mathsf{V}}^\prime = 419$}\\
\bottomrule
\end{tabular}
\end{adjustbox}
\\\emph{Note:} $N_{\mathsf{B}} = 379$, $N_{\mathsf{V}} = 433$ are after attention checks.
\end{table}

\subsection{Descriptives}
\label{sec:desc}

Evaluating the sample distribution, we found the indicator variables to be \refterm{skewness}{negatively skewed}. The distributions tail off to the left. The Control and Awareness indicators suffer from positive kurtosis. We found that the indicator distributions as well as the IUIPC sub-scale distributions exhibited substantial \refterm{normal_distribution}{non-normality}. We illustrate these aspects in Table~\ref{tab:descSubScalesRedux} and Figure~\ref{fig:densitySubScalesRedux}. 

\descSubScalesRedux
\densitySubScalesRedux

We observed that the two samples displayed approximately equal distributions by sub-scales. 
Controlling for the difference between Samples \textsf{B} and \textsf{V}, we found that none of their sub-scale means were statistically significantly different, the maximal absolute standardized mean difference being 0.13---a small magnitude.

Our IUIPC-10 samples yielded $6\%$ \refterm{outlier}{univariate outliers} by the robust outlier labeling rule and $3\%$ multi-variate outliers with a Mahalanobis distance of $12$ or greater~\cite[pp. 72]{kline2015principles}. We removed these MV outliers as indicated in Table~\ref{tab:sample}.

These observations on the distribution of the samples are relevant for the choice of estimator for the confirmatory factor analysis to come. An Maximum Likelihood (ML) estimation would require \refterm{continuous}{continuous measurement} with \refterm{normal_distribution}{multi-variate normality}. These assumptions are clearly not fulfilled. While a robust Maximum Likelihood estimation (MLM) can also be considered, we opted for a Diagonally Weighted Least Square (DWLS) estimation with robust standard errors and scale-shifted mean- and variance-adjusted test statistics (WLSMV), typically considered preferred for ordinal/non-normal data.\footnote{Gro{\ss}~\cite{Gross2021IUIPC} also employed a DWLS-estimation in the evaluation of IUIPC-8. The models of that work, however, were computed with WLSMVS, an estimation method using Satterthwaite-style test statistic scaling.}

\subsection{Construct Validity}
\label{sec:construct_val}

\subsubsection{Factorial Validity}
\label{sec:factor_val}


\begin{table*}[tbp]
\centering
\caption{Comparison of different model structures of IUIPC-8 on Sample \textsf{B} with WLSMV estimation}
\label{tab:comparison.models.wls}
\begin{tabular}{@{}rrrrrrrrrrrrr@{}}
\toprule
& \multicolumn{3}{c}{One Factor}& \multicolumn{3}{c}{Two Factors}& \multicolumn{3}{c}{Three Factors (1\textsuperscript{st} Order)}& \multicolumn{3}{c}{Three Factors (2\textsuperscript{nd} Order)}\tabularnewline 
\midrule
$\chi^{2} (\mathit{df})$ & 732.43 (20)& & & 88.98 (19) & & & 22.77 (17)& &  & 22.77 (17)& & \tabularnewline
$\chi^2/\mathit{df}$ & 36.62 &&& 4.68 &&& 1.34 &&& 1.34 \tabularnewline
\textsf{CFI} & .93& & & .99& & & 1.00& & & 1.00& & \tabularnewline
\textsf{RMSEA} & .32& \multicolumn{2}{c}{[.30, .34]} & .11& \multicolumn{2}{c}{[.09, .13]} & .07& \multicolumn{2}{c}{[.05, .09]} & .07 & \multicolumn{2}{c}{[.05, .09]} \tabularnewline
\textsf{SRMR}& .22& & & .09& & & .04 & & & .04& & \tabularnewline
Scaled $\chi^{2} (\mathit{df})$& 790.34 (20)& & & 105.24 (19)& & & 46.76 (17)& & & 46.76 (17) & & \tabularnewline
\bottomrule
\end{tabular}
\end{table*}

First, we investigate the three-dimensionality of IUIPC-8. To that end, we computed confirmatory factor analyses on one-factor, two-factor and the hypothesized three-dimensional second-order model displayed in Table~\ref{tab:comparison.models.wls}. 
The two-factor solution was statistically significantly better than the one-factor solution, $\chi^2(1) = 215.065, p < .001$. In turn, the three-factor solutions were statistically significantly better than the two-factor solution, $\chi^2(2) = 30.165, p < .001$. Hence, given the results of the Likelihood-Ratio tests (LRTs) on these nested models, we choose the three-dimensional second-order model. This is the model also shown in the path plot of Figure~\ref{fig:pathPlotCFABreduxwlsref}.

\subsubsection{Model Fit}
\paragraph*{Global Fit}

Second, we evaluate the global fit as a measure of factorial validity. We do this in a two-way comparison of WLSMV CFAs on the following dimensions:
\begin{inparaenum}[(i)]
  \item IUIPC-10 vs. IUIPC-8, and
  \item base sample \textsf{B} and validation sample \textsf{V}.
\end{inparaenum}
Table~\ref{tab:respec} reports on the fit statistics of the four CFA models.

\begin{table*}
\centering
\caption{Fit statistic comparison of IUIPC-10 and IUIPC-8}
\label{tab:respec}
\begin{adjustbox}{max width=\textwidth}
\begin{tabular}{lcc}
\toprule
\multirow{2}{*}{Instrument} & \multicolumn{2}{c}{Sample}\\
\cmidrule(lr){2-3}
& Base \textsf{B} & Validation \textsf{V} \\
\midrule
\multirow{4}{*}{IUIPC-10} & $\chi^2(32) = 275.087, p < .001$ & $\chi^2(32) = 220.96, p < .001$\\ 
 & \textsf{CFI}$^\ddagger$=$.96$ & 
    \textsf{CFI}=$.96$\\
 & \textsf{RMSEA}$^\ddagger$=$.14$ $[.13, .16]$, $p_{\epsilon_0 \leq .05} < .001$ & \textsf{RMSEA}$^\ddagger$=$.12$ $[.10, .13]$, $p_{\epsilon_0 \leq .05} < .001$ \\
 & \textsf{SRMR}=$.10$ &
    \textsf{SRMR}=$.07$\\
\cmidrule(lr){2-2}\cmidrule(lr){3-3}
\multirow{4}{*}{IUIPC-8}                  & $\chi^2(17) = 46.764, p < .001$ &
$\chi^2(17) = 36.673, p = .004$\\
 & \textsf{CFI}$^\ddagger$=$1.00$ &
    \textsf{CFI}$^\ddagger$=$1.00$ \\
 & \textsf{RMSEA}$^\ddagger$=$.07$ $[.05, .09]$, $p_{\epsilon_0 \leq .05} = .086$ & 
    \textsf{RMSEA}$^\ddagger$=$.05$ $[.03, .08]$, $p_{\epsilon_0 \leq .05} = .394$ \\
 & \textsf{SRMR}=$.04$ &
    \textsf{SRMR}=$.03$\\
\bottomrule
\end{tabular}
\end{adjustbox}
\\\emph{Note:} $^\ddagger$ Robust estimation with scaled test statistic. \textsf{RMSEA} reported wih 90\% CI.
\end{table*}

Because IUIPC-10 and IUIPC-8 models are non-nested, we cannot use Likelihood Ratio test (LRT) to evaluate their difference.  In the WLSMV estimation, we are left with comparing fit measures.\footnote{On the Maximum Likelihood (ML) estimation used by Gro{\ss}~\cite{Gross2021IUIPC}, a Vuong non-nested LRT was available as a test of difference. That work also considered a Consistent Akaike Information Criterion (CAIC) directly derived from the $\chi^2$ metric typically used for ML estimations. Banks and Joyner~\cite{banks2017aic} offer detailed analysis of AICs for different estimations.}

Regarding the global fit reported in Table~\ref{tab:respec}, we notice that all CFA models fail the exact-fit test on the $\chi^2$ test statistic. To evaluate approximate fit, we draw attention to the Root Mean Square Estimate of Approximation (RMSEA), its confidence interval and the close-fit hypothesis $H_{\varepsilon_0 \leq .05, 0}$. We observe that the IUIPC-10 models are not supported by the close-fit test, the IUIPC-8 models are, Sample \textsf{B}: $p_{\epsilon_0 \leq .05} = .086$; Sample \textsf{V}: $p_{\epsilon_0 \leq .05} = .394$.
Hence, we conclude that the IUIPC-8 model shows a sufficient approximate close fit, even if not an exact fit.

\paragraph*{Local Fit}
\label{sec:local.fit}

The good global fit for IUIPC-8 shown in Table~\ref{tab:respec} alone is not sufficient to vouch for the overall fit of the model. 
For this purpose, we inspect the correlation and raw residuals in Table~\ref{tab:residualsBreduxwls}. Therein, we observe slightly reduced correlation residuals between \textsf{coll1} and the Awareness indicator variables. These negative correlation residuals mean that the CFA model overestimates the correlation between the indicator variables in question. The correlation residuals in the validation model \processifversion{DVSupplementaryAppendix}{(Table~\ref{tab:residualsVreduxwls} in Appendix~\ref{ref:validation})}\processifversion{DVSupplementaryExternal}{(included in the online supplementary materials)} show lower deviations. Hence, we believe both models to hold acceptable local fit.

\residualsBreduxwls

\subsubsection{CFA Model, Convergent and Discriminant Validity}

We illustrate the selected second-order CFA model for IUIPC-8 in Figure~\ref{fig:pathPlotCFABreduxwlsref}. Table~\ref{tab:loadingsBreduxwls} contains the corresponding factor loadings with their standardized solutions. The standardized loadings of the model give us confidence in the \refterm{convergent_validity}{convergent validity} of the model: the Average Variance Extracted (AVE) was greater than $50\%$ for all first-level factors. This observation holds equally for the standardized factor loadings of the validation CFA\processifversion{DVSupplementaryAppendix}{, summarized in Table~\ref{tab:loadingsVreduxwls} in Appendix~\ref{ref:validation}}\processifversion{DVSupplementaryExternal}{, summarized in the online supplementary materials}.

\newcommand{\pathPlotCFABreduxwlsref}{
\begin{figure*}[tbp]
\centering
\includegraphics[keepaspectratio,width=0.8\maxwidth]{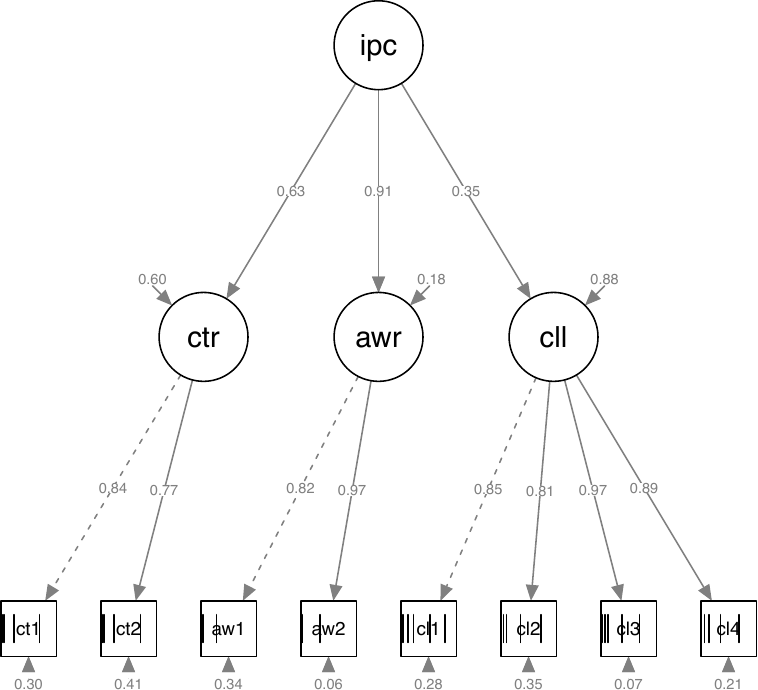} 
\caption{CFA paths plot with standardized estimates of IUIPC-8 on Sample \textsf{B}. \emph{Note:} The dashed lines signify that the raw factor loading was fixed to 1 (cf. Table~\ref{tab:loadingsBreduxwls}).}
\label{fig:pathPlotCFABreduxwlsref}
\Description[The path plot of the IUIPC-8 model on the base sample B showing the three first-level factors ctrl, aware, and collect and the second-level factor iuipc.]{This path plot shows the final IUIPC-8 model on base sample B. On the bottom, we have eight indicator variables, for which vertical lines depict the thresholds of the WLSMV model. In the middle, we have the three first-level factors ctrl, aware, and collect. On the top, we show the second-level factor iuipc.}
\end{figure*}
}
\pathPlotCFABreduxwlsref

\loadingsBreduxwls

In terms of \refterm{discriminant_validity}{discriminant validity}, we first consider the Fornell-Larcker criterion in Table~\ref{tab:corfitBVredux}. As required, we find that the square-root of the AVE displayed on the diagonal of the matrices is greater than the inter-factor correlations in the rest of the matrix. This holds for both the base Sample \textsf{B} and the validation Sample \textsf{V}.
\corfitBVredux

Further, we evaluate the \refterm{heterotrait_monotrait_ratio}{HTMT} criterion in Table~\ref{tab:htmtBVredux}. We are satisfied with this criterion for Samples~\textsf{B} and~\textsf{V} at a threshold of $.85$. Hence, we conclude that the scale offers sufficient \refterm{discriminant_validity}{discriminant validity}.
\htmtBVredux

\subsection{Reliability: Internal Consistency}

Table~\ref{tab:loadingsBreduxwls} also includes reliability metrics derived from the WLSMV CFA model of IUIPC-8. For both the base Sample~\textsf{B} and the validation Sample~\textsf{V}\processifversion{DVSupplementaryAppendix}{ (Table~\ref{tab:loadingsVreduxwls} in Appendix~\ref{ref:validation})}, we observe that the congeneric reliability $\omega$ is consistently greater than $.70$. By that, the reliability criteria established in Section~\ref{sec:criteria} are fulfilled and we can expect a good signal-to-noise ratio for the scale.


\newcommand{\SUCC}{\CIRCLE}
\newcommand{\PART}{\LEFTcircle}
\newcommand{\FAIL}{\Circle}

\begin{table*}
\centering
\footnotesize
\caption{Selected evidence for construct validity and reliability criteria on Samples \textsf{B} and \textsf{V} under WLSMV estimation}
\label{tab:evidence}
\begin{tabular}{llcccccccccc}
\toprule
& & \multicolumn{8}{c}{Construct Validity} & \multicolumn{2}{c}{Reliability}\tabularnewline
\cmidrule(lr){3-10}\cmidrule(lr){11-12}
&& \multicolumn{4}{c}{Factorial} & \multicolumn{2}{c}{Convergent} & \multicolumn{2}{c}{Divergent} & \multicolumn{2}{c}{Internal Consistency}\tabularnewline
\cmidrule(lr){3-6} \cmidrule(lr){7-8} \cmidrule(lr){9-10} \cmidrule(lr){11-12}
&& $H_{\chi^2, 0}$ & $H_{\hat{\varepsilon} \leq .05, 0}$ & $H_{\hat{\varepsilon} > .05, 0}$ & $H_{\hat{\varepsilon} > .10, 0}$ & $\beta > .70$ & $\vari{AVE} > .50$ & $\sqrt{AVE} > \forall\bar{r}$ & $\vari{HTMT} < .85$ & $\alpha > .70$  & $\omega > .70$ \tabularnewline
\midrule
\multirow{2}{*}{IUIPC-10} 
  & \textsf{B} &  \FAIL & \FAIL & \FAIL & \FAIL & \FAIL & \PART & \FAIL & \SUCC & \FAIL & \FAIL \tabularnewline
  & \textsf{V} & \FAIL &  \FAIL & \FAIL & \FAIL & \FAIL & \FAIL & \FAIL & \SUCC & \FAIL & \FAIL \tabularnewline
\midrule
\multirow{2}{*}{IUIPC-8}
  & \textsf{B} & \FAIL & \SUCC & \FAIL & \SUCC & \SUCC & \SUCC & \SUCC & \SUCC & \SUCC & \SUCC \tabularnewline
  & \textsf{V} & \FAIL & \SUCC & \FAIL & \SUCC & \SUCC & \SUCC & \SUCC & \SUCC & \SUCC & \SUCC \tabularnewline
\bottomrule
\end{tabular}
\\\emph{Note:} $\SUCC$ = supports the model; $\FAIL$ = rejects the model;
$\beta$ = standardized loading; \vari{AVE} = Average Variance Extracted; $\bar{r}$ = correlation with other factor; \vari{HTMT} = Heterotrait-Monotrait Ratio; $\omega$ = Composite Reliability.
\end{table*}     

\section{Discussion}
\label{sec:discussion}

We have seen that IUIPC-8 offers good construct validity and reliability. The outcomes of our analysis are summarized in Table~\ref{tab:evidence}.
It can serve as a useful measurement instrument for information privacy concern. As any measurement instrument, IUIPC-8 is a working solution, which may be proven wrong eventually and superseded by more refined scales~\cite{john2000measurement}.  

In terms of bandwidth-fidelity trade-off~\cite{john2000measurement}, IUIPC-8 offers a greater fidelity than the original scale IUIPC-10~\cite{malhotra2004internet}, at the expense of bandwidth. Because of the greater congeneric reliability, we expect less attenuation in the correlations to other variables that with IUIPC-10: according to classical test theory, such correlations are bounded by the square-root of its reliability. Thereby, IUIPC-8 can be useful in investigating relations to other variables, such as impact of privacy concern on privacy behavior.

The restriction to eight items also bears limitations that need to be considered carefully. First, the factors Control and Awareness are based on a narrower footing in terms of content validity, that is, in terms of coverage of relevant and representative aspects of the construct. We also need to consider CFA model identification. While IUIPC-8 as a whole is identified because of the two-indicator rule~\cite[Rule 9.1]{kline2015principles}, the sub-scales Control and Awareness on their own will not be identified. Hence, they cannot be used as robust measurement instruments of their own.

In terms of future work, it would be preferable to refine IUIPC-8 with further items rounding out the sub-scales Control and Awareness, while maintaining a high construct validity and reliability. Ideally, each factor would have three or more indicators. 
While this chapter largely focused on the construct validity and reliability in the form of internal consistency of the scale itself, a more comprehensive evaluation of privacy concern scales is vital. For IUIPC-8 we considered internal consistency as reliability (that is, generalizing across items). At the same time, retest reliability (generalizability across times) and equivalence reliability (generalizability across forms) are still research areas to expand. In addition, the investigation of IUIPC in its full nomology are important, such as pursued by Pape et al.~\cite{pape2020re} in the case of the use of the scale in Japan.

\section{SUMMARY}
This chapter considers the validity and reliability of privacy concern scales, with IUIPC-8 as an example of a brief information privacy concern scale.
\begin{compactitem}[$\blacksquare$]
\item We introduced validity and reliability concepts, focusing on construct validity and congeneric reliability.
\item We discussed confirmatory factory analysis as a tool to establish the properties of measurement instruments.
\item We discussed CFA estimation methods for ordinal and non-normal data as found with the IUIPC scale.
\item We included an empirical analysis of IUIPC-8 on both a base sample and an independent validation sample.
\item We evaluated validity and reliability criteria in the comparison of IUIPC-10 and the brief form IUIPC-8.
\end{compactitem}

\section*{Acknowledgments}
This work was supported by \CASCAde.

\bibliographystyle{ACM-Reference-Format}
\bibliography{./sem,./privacy,./methods_resources,./qdesign,./pets_use_IUIPC,./validity,./reliability,./mturk,iuipc_analysis}

\begin{appendix}

\section{Materials \& Sample}
\label{app:sample}
\label{app:materials}

We included the used IUIPC-10 questionnaire in Table~\ref{tab:iuipc}.
\begin{table*}[tb]
\centering
\caption{Items of the instrument Internet users' information privacy concerns (IUIPC-10)~\cite{malhotra2004internet}, adapted from~\cite{Gross2021IUIPC}}
\label{tab:iuipc}
\begin{tabular}{lcp{.7\textwidth}}
  \toprule
Construct & Item & Question\\ 
  \midrule
\multirow{3}{*}{Control (\fVar{ctrl})} & \iVar{ctrl1} & Consumer online privacy is really a matter of consumers' right to exercise control and autonomy over decisions about how their information is collected, used, and shared.\\
 & \iVar{ctrl2} & Consumer control of personal information lies at the heart of consumer privacy.\\
 & [\iVar{ctrl3}] & I believe that online privacy is invaded when control is lost or unwillingly reduced as a result of a marketing transaction.\\
\addlinespace
 \multirow{3}{*}{Awareness (\fVar{aware})} & \iVar{awa1} & Companies seeking information online should disclose the way the data are collected, processed, and used.\\
 & \iVar{awa2} & A good consumer online privacy policy should have a clear and conspicuous disclosure.\\
 & [\iVar{awa3}] & It is very important to me that I am aware and knowledgeable about how my personal information will be used.\\
 \addlinespace
 \multirow{4}{*}{Collection (\fVar{collect})} & \iVar{coll1} & It usually bothers me when online companies ask me for personal information.\\
 & \iVar{coll2} & When online companies ask me for personal information, I sometimes think twice before providing it.\\
 & \iVar{coll3} & It bothers me to give personal information to so many online companies.\\
 & \iVar{coll4} & I'm concerned that online companies are collecting too much personal information about me.\\
\bottomrule
\end{tabular}
\\\emph{Note:} The questionnaire is administered with 7-point Likert items, anchored on 1=``Strongly Disagree'' to 7=``Strongly Agree.'' The items in squared brackets \iVar{ctrl3} and \iVar{awa3} are included in IUIPC-10, but not in IUIPC-8.
\end{table*}

For the reproducibility of the maximum likelihood estimation, Table~\ref{tab:inputCorSDMB} contains the correlations, means and standard deviations (SDs) of Sample~\textsf{B}. The OSF supplementary materials contain more precise covariance matrices of all samples.

\inputCorSDMB

\section{Thresholds}
In this section, we include the WLSVM thresholds for the base and the validation models. Table~\ref{tab:thresholdsB} shows the thresholds of the main model.

\thresholdsB

\begin{DVSupplementaryAppendix}
\section{Validation CFA}
\label{ref:validation}

In this section, we include the evaluation of the IUIPC-8 CFA on the validation sample \textsf{V}.
Table~\ref{tab:inputCorSDMV} includes the sample correlations, means and standard deviations of \textsf{V}.

\inputCorSDMV

We are showing the progression of one-factor, two-factor and three-factor model on Sample~\textsf{V} in Table~\ref{tab:comparison.models.wls.V}. We display the chosen model in the path model in Figure~\ref{fig:pathPlotCFAVreduxwlsref}.

\begin{table*}[tbp]
\centering
\caption{Comparison of different model structures of IUIPC-8 on Sample \textsf{V} with WLSMV estimation}
\label{tab:comparison.models.wls.V}
\begin{tabular}{@{}rrrrrrrrrrrrr@{}}
\toprule
& \multicolumn{3}{c}{One Factor}& \multicolumn{3}{c}{Two Factors}& \multicolumn{3}{c}{Three Factors (1\textsuperscript{st} order)}& \multicolumn{3}{c}{Three Factors (2\textsuperscript{nd} Order)}\tabularnewline
\midrule
$\chi^{2}(\mathit{df})$ & 814.81 (20)& & & 105.01 (19)& & & 18.81 (17)& & & 18.81 (17)& & \tabularnewline
$\chi^2/\mathit{df}$ &             40.74 & & &         5.53    & & &  1.11         & & &  1.11\tabularnewline
\textsf{CFI} & .90& & & .99& & & 1.00& & & 1.00& & \tabularnewline
\textsf{RMSEA} & .30& \multicolumn{2}{c}{[.29, .32]} & .12& \multicolumn{2}{c}{[.10, .14]} & .05& \multicolumn{2}{c}{[.03, .08]} & 0.05&  \multicolumn{2}{c}{[.03, .08]} \tabularnewline
\textsf{SRMR} & .19& & & .08& & & .03& & & .03& & \tabularnewline
Scaled $\chi^{2}(\mathit{df})$& 794.27 (20)& & & 133.63 (19)& & &  36.67 (17)& & & 36.67 (17)& & \tabularnewline
\bottomrule
\end{tabular}
\end{table*}

\pathPlotCFAVreduxwlsref

\paragraph*{Global Fit} 
The global fit of the validation model was already inspected in Table~\ref{tab:respec} in Section~\ref{sec:factor_val}.

\paragraph*{Local Fit}

The residuals displayed in Table~\ref{tab:residualsVreduxwls} are all in good shape, indicating a good local fit throughout. 
\residualsVreduxwls

\paragraph*{CFA Model, Convergent and Discriminant Validity}

Considering the standardized loadings in Table~\ref{tab:loadingsVreduxwls}, we find that the indicator variables explain more consistently more than $50\%$ of the variance. 

\loadingsVreduxwls

\paragraph*{Threshold}
Table~\ref{tab:thresholdsV} the corresponding thresholds for the validation.
\thresholdsV
\end{DVSupplementaryAppendix}
\end{appendix}

\end{document}